\documentclass[prd,aps,groupedaddress,showpacs]{revtex4}
\usepackage{epsf,epsfig,graphicx}
\newcounter{muni}

\begin{document}
\hbadness=10000 \pagenumbering{arabic}
\rightline{DPNU-05-09}
\rightline{TU-748}

\title{Resolution to the $B\to \pi K$ puzzle}

\author{Hsiang-nan Li$^{1}$}
\email{hnli@phys.sinica.edu.tw}
\author{Satoshi Mishima$^2$}
\email{mishima@tuhep.phys.tohoku.ac.jp}
\author{A.I. Sanda$^3$}
\email{sanda@eken.phys.nagoya-u.ac.jp}

\affiliation{$^{1}$Institute of Physics, Academia Sinica, Taipei,
Taiwan 115, Republic of China}
\affiliation{$^{1}$Department of Physics, National Cheng-Kung University,\\
Tainan, Taiwan 701, Republic of China}
\affiliation{$^{2}$Department of Physics, Tohoku University,
Sendai 980-8578, Japan} \affiliation{$^{3}$Department of Physics,
Nagoya University, Nagoya 464-8602, Japan}

\begin{abstract}

We calculate the important next-to-leading-order contributions to
the $B\to \pi K$, $\pi\pi$ decays from the vertex corrections, the
quark loops, and the magnetic penguins in the perturbative QCD
approach. It is found that the latter two reduce the leading-order
penguin amplitudes by about 10\%, and modify only the $B\to\pi K$
branching ratios. The main effect of the vertex corrections is to
increase the small color-suppressed tree amplitude by a factor of
$3$, which then resolves the large difference between the direct
CP asymmetries of the $B^0\to \pi^\mp K^\pm$ and $B^\pm\to \pi^0
K^\pm$ modes. The puzzle from the large $B^0\to\pi^0\pi^0$
branching ratio still remains.

\end{abstract}


\maketitle

\section{INTRODUCTION}

The $B$ factories have accumulated enough events, which allow
precision measurements of exclusive $B$ meson decays. These
measurements sharpen the discrepancies between experimental data
and theoretical predictions within the standard model, such that
some puzzles have appeared. The recently observed direct CP
asymmetries and branching ratios of the $B\to \pi K$, $\pi\pi$
decays \cite{HFAG},
\begin{eqnarray}
A_{CP}(B^0\to \pi^\mp K^\pm)&=&(-11.5\pm 1.8)\%\;,\nonumber\\
A_{CP}(B^\pm\to \pi^0 K^\pm)&=&(4\pm 4)\%\;,\nonumber\\
B(B^0\to\pi^+\pi^-)&=&(5.0\pm 0.4)\times 10^{-6}\;,\nonumber\\
B(B^0\to\pi^0\pi^0)&=&(1.45\pm 0.29)\times 10^{-6}\;,\label{data}
\end{eqnarray}
are prominent examples. The expected relations $A_{CP}(B^0\to
\pi^\mp K^\pm)\approx A_{CP}(B^\pm\to \pi^0 K^\pm)$ and
$B(B^0\to\pi^+\pi^-)\gg B(B^0\to\pi^0\pi^0)$ obviously contradict
to the above data. In this work we shall investigate the
indication of Eq.~(\ref{data}), and study whether they can be
accommodated in the perturbative QCD approach \cite{KLS,LUY}.

To explain these puzzles, it is useful to adopt the
topological-amplitude parameterization \cite{CC} for these decays.
The most general parameterization of the $B\to\pi\pi$ decay
amplitudes is written as
\begin{eqnarray}
A(B^0\to \pi^+\pi^-)&=&-T\left(1 +\frac{P}{T}e^{i\phi_2}\right)\;,
\nonumber\\
\sqrt{2}A(B^+\to \pi^+\pi^0)&=&-T\left[1+\frac{C}{T}
+\frac{P_{ew}}{T}e^{i\phi_2}\right]\;,
\nonumber\\
\sqrt{2}A(B^0\to \pi^0\pi^0)&=&T\left[\left(
\frac{P}{T}-\frac{P_{ew}}{T}\right)
e^{i\phi_2}-\frac{C}{T}\right]\;, \label{Mbpi1}
\end{eqnarray}
where $T$, $C$, $P$, and $P_{ew}$ stand for the color-allowed
tree, color-suppressed tree, penguin, and electroweak penguin
amplitudes, respectively, and $\phi_2$ is the weak phase defined
later. The counting rules in terms of powers of the Wolfenstein
parameter $\lambda\sim 0.22$ are then assigned to various decay
amplitudes \cite{GHL,Charng,Y03}. The amplitudes in
Eq.~(\ref{Mbpi1}) obey the counting rules in the standard model,
\begin{eqnarray}
& &\frac{P}{T}\sim \lambda\;,\;\;\;\; \frac{C}{T}\sim \lambda\;,
\;\;\;\; \frac{P_{ew}}{T}\sim \lambda^2\;. \label{po}
\end{eqnarray}
Therefore, the $B^0\to \pi^0\pi^0$ branching ratio is expected to
be of $O(\lambda^2)$ of the $B^0\to \pi^+\pi^-$ one. However,
Eq.~(\ref{data}) shows that the former is about of $O(\lambda)$ of
the latter.

The $B\to \pi K$ decay amplitudes are written, up to
$O(\lambda^2)$, as
\begin{eqnarray}
A(B^+\to \pi^+ K^0)&=&P'\;,\nonumber\\
\sqrt{2}A(B^+\to \pi^0 K^+)&=&-P'\left[1+\frac{P'_{ew}}{P'}
+\left(\frac{T'}{P'}+\frac{C'}{P'} \right)e^{i\phi_3}\right]\;,
\nonumber\\
A(B^0\to \pi^-
K^+)&=&-P'\left(1+\frac{T'}{P'}e^{i\phi_3}\right)\;,
\nonumber\\
\sqrt{2}A(B^0\to \pi^0K^0)&=&P'\left(1 -\frac{P'_{ew}}{P'}
-\frac{C'}{P'}e^{i\phi_3}\right)\;, \label{Mbpp1}
\end{eqnarray}
where the notations $T'$ $C'$, $P'$ and $P_{ew}'$ bear the
meanings the same as for the $B\to\pi\pi$ decays but with primes,
and the weak phase $\phi_3$ is defined via the
Cabibbo-Kobayashi-Maskawa (CKM) matrix element
$V_{ub}=|V_{ub}|\exp(-i\phi_3)$ \cite{KM}. These amplitudes obey
the counting rules,
\begin{eqnarray}
\frac{T'}{P'}\sim  \lambda\;,\;\;\;\;
\frac{P'_{ew}}{P'}\sim\lambda\;,\;\;\;\;
\frac{C'}{P'}\sim\lambda^2\;. \label{pow}
\end{eqnarray}
The data $A_{CP}(B^0\to \pi^\mp K^\pm)\approx -11\%$ implies a
sizable relative strong phase between $T'$ and $P'$, which
verifies our prediction made years ago using the PQCD approach
\cite{KLS}. Since both $P_{ew}'$ and $C'$ are subdominant, the
approximate equality for the direct CP asymmetries
$A_{CP}(B^\pm\to \pi^0K^\pm)\approx A_{CP}(B^0\to \pi^\mp K^\pm)$
is expected, which is, however, in conflict with the data in
Eq.~(\ref{data}) dramatically.

It is then natural to conjecture a large $P_{ew}'$
\cite{Y03,BFRPR,NK04,GR03,CFMM}, which signals a new physics
effect, a large $C'$ \cite{Charng2,HM04,CGRS,Ligeti04}, or both
\cite{WZ,BHLD} in the $B\to\pi K$ decays. The large $C'$ proposal
seems to be favored by a recent analysis of the $B\to \pi K$,
$\pi\pi$ data based on the amplitude parameterization
\cite{Charng2}. Note that the current PQCD predictions for the
two-body nonleptonic $B$ decays were derived from the
leading-order (LO) formalism. While LO PQCD implies a negligible
$C'$, it is possible that this supposedly tiny amplitude may
receive a significant subleading correction. Hence, before
claiming a new physics signal, one should at least examine whether
the next-to-leading-order (NLO) effects could enhance $C'$
sufficiently. In this paper we shall calculate the important NLO
contributions to the $B\to \pi K$, $\pi\pi$ decays from the vertex
corrections, the quark loops, and the magnetic penguins. The
higher-power corrections have not yet been under good control, and
will not be considered here. We find that the corrections from the
quark loops and from the magnetic penguins, being about 10\% of
the LO penguin amplitude, decrease only the $B\to\pi K$ branching
ratios. The vertex corrections tend to increase $C'$ by a factor
of 3. This larger $C'$ leads to nearly vanishing $A_{CP}(B^\pm\to
\pi^0K^\pm)$ without changing the branching ratios, which are
governed by $P'$. The $B\to \pi K$ puzzle is then resolved.

The other NLO corrections, mainly to the $B$ meson transition form
factors, can be eliminated by choosing an appropriate
renormalization scale $\mu\sim \sqrt{\bar\Lambda m_b}$,
$\bar\Lambda$ being a hadronic scale and $m_b$ the $b$ quark mass.
This observation follows the well-known Brodsky-Lepage-Mackenzie
(BLM) procedure \cite{BLM}, in which the scale $\mu$ is determined
in the way that the vacuum polarization effects are absorbed into
the coupling constant $\alpha_s(\mu)$. It has been demonstrated
with this procedure that NLO corrections to many exclusive
processes are minimized to some extent \cite{BLM}. Taking the
simple pion form factor as an example, the BLM scale has been
found to be of the order of the invariant mass of the hard
exchanged gluon. The choice of $\mu$ proposed in the PQCD approach
\cite{LS} is basically in agreement with this procedure: the
argument $\mu$ of the coupling constant is set to the invariant
masses of internal particles, which are of $O(\sqrt{\bar\Lambda
m_b})$ for the $B$ meson transition form factors
\cite{TLS,CKL,Mishima:2001ms}. A general feature of the BLM scale
is that it is always much lower than the external kinematic
variable, implying that the smallness of the coupling constant is
not the only condition for the applicability of perturbation
theory.

As mentioned before, the observed branching ratio
$B(B^0\to\pi^0\pi^0)\approx 1.5\times 10^{-6}$ is much larger than
the LO PQCD prediction $\sim 10^{-7}$ \cite{LUY,KS}. The
prediction from QCD-improved factorization (QCDF) has the same
order of magnitude \cite{BBNS}. Since this mode involves a
subdominant color-suppressed tree amplitude as shown in
Eq.~(\ref{Mbpi1}), a larger $C$ certainly helps to resolve the
$B\to\pi\pi$ puzzle. We also compute the NLO corrections to these
decays, and find the similar reduction from the quark loops and
the magnetic penguins, which are about 10\% of the LO penguin
amplitude $P$. Since $P$ is subdominant, the
$B^0\to\pi^\mp\pi^\pm$ and $B^\pm\to\pi^\pm\pi^0$ branching ratios
almost remain the same. The enhancement of $C$ from the vertex
corrections, leading to $B(B^0\to\pi^0\pi^0)\approx 0.3\times
10^{-6}$, is still not sufficient to account for the data. We
point out that any new mechanism, introduced to resolve this
puzzle, must survive the constraint from the tiny observed
branching ratios \cite{HFAG},
\begin{eqnarray}
B(B^0\to K^0\overline K^0)&=& (0.96^{+0.25}_{-0.24})\times 10^{-6}\;,
\nonumber\\
B(B^0\to \rho^0\rho^0)&<& 1.1\times 10^{-6}\;.
\end{eqnarray}
The leading PQCD predictions for $B(B^0\to K^0\overline K^0)$
\cite{CL01} and for $B(B^0\to\rho^0\rho^0)$ \cite{LILU,rho} have
been consistent with the experimental data. The proposals of
final-state interaction \cite{CCS} and of the charming penguin in
soft-collinear effective theory (SCET) \cite{BPRS} have not yet
been applied to the $B^0\to\rho^0\rho^0$ decay.

We review the LO PQCD predictions for the $B\to \pi K$, $\pi\pi$
decays, including those for the mixing-induced CP asymmetries, in
Sec.~II. The vertex corrections, the quark loops, and the magnetic
penguin amplitudes are computed in Sec.~III. We perform the
numerical study in Sec.~IV, where the theoretical uncertainty is
also analyzed. Section V is the conclusion. The explicit
factorization formulas for the various topologies of decay
amplitudes are collected in Appendix.

\section{LEADING-ORDER PREDICTIONS}

The effective Hamiltonian for the $b\to s$ transition is given by
\cite{REVIEW},
\begin{equation}
H_{\rm eff}\, =\, {G_F\over\sqrt{2}}
\sum_{q=u,c}V_{qb}V^*_{qs}\left[C_1(\mu)O_1^{(q)}(\mu)
+C_2(\mu)O_2^{(q)}(\mu)+ \sum_{i=3}^{10}C_i(\mu)O_i(\mu)\right]\;,
\label{hbk}
\end{equation}
with the Fermi constant $G_F=1.16639\times 10^{-5}\;{\rm
GeV}^{-2}$, and the CKM matrix elements $V$. The four-fermion
operators are written as
\begin{eqnarray}
{\renewcommand\arraystretch{1.5}
\begin{array}{ll}
\displaystyle
O_1^{(q)}\, =\,
(\bar{s}_iq_j)_{V-A}(\bar{q}_jb_i)_{V-A}\;,
& \displaystyle
O_2^{(q)}\, =\, (\bar{s}_iq_i)_{V-A}(\bar{q}_jb_j)_{V-A}\;,
\\
\displaystyle
O_3\, =\, (\bar{s}_ib_i)_{V-A}\sum_{q'}(\bar{q}'_jq'_j)_{V-A}\;,
& \displaystyle
O_4\, =\, (\bar{s}_ib_j)_{V-A}\sum_{q'}(\bar{q}'_jq'_i)_{V-A}\;,
\\
\displaystyle
O_5\, =\, (\bar{s}_ib_i)_{V-A}\sum_{q'}(\bar{q}'_jq'_j)_{V+A}\;,
& \displaystyle
O_6\, =\, (\bar{s}_ib_j)_{V-A}\sum_{q'}(\bar{q}'_jq'_i)_{V+A}\;,
\\
\displaystyle
O_7\, =\,
\frac{3}{2}(\bar{s}_ib_i)_{V-A}\sum_{q'}e_{q'}(\bar{q}'_jq'_j)_{V+A}\;,
& \displaystyle
O_8\, =\,
\frac{3}{2}(\bar{s}_ib_j)_{V-A}\sum_{q'}e_{q'}(\bar{q}'_jq'_i)_{V+A}\;,
\\
\displaystyle
O_9\, =\,
\frac{3}{2}(\bar{s}_ib_i)_{V-A}\sum_{q'}e_{q'}(\bar{q}'_jq'_j)_{V-A}\;,
& \displaystyle
O_{10}\, =\,
\frac{3}{2}(\bar{s}_ib_j)_{V-A}\sum_{q'}e_{q'}(\bar{q}'_jq'_i)_{V-A}\;,
\end{array}
}
\label{o}
\end{eqnarray}
with the color indices $i, \ j$, and the notations
$(\bar{q}'q')_{V\pm A} = \bar q' \gamma_\mu (1\pm \gamma_5)q'$.
The index $q'$ in the summation of the above operators runs
through $u,\;d,\;s$, $c$, and $b$. The effective Hamiltonian for
the $b\to d$ transition is obtained by changing $s$ into $d$ in
Eqs.~(\ref{hbk}) and (\ref{o}).

According to Eq.~(\ref{hbk}), the amplitude for a $B$ meson decay
into the final state $f$ through the $\bar b\to \bar s(\bar d)$
transition has the general expression,
\begin{eqnarray}
{\cal A}(B \to f) &=& V_{ub}^*V_{us(d)}\, {\cal A}^{(u)}_{f}
 +V_{cb}^*V_{cs(d)}\, {\cal A}^{(c)}_{f}
 +V_{tb}^*V_{ts(d)}\, {\cal A}^{(t)}_{f}\;.
%\nonumber\\
%&=&
% V_{ub}^*V_{uq} \left(A^{(u)}_{f} -A^{(t)}_{f}  \right)
% +V_{cb}^*V_{cq} \left(A^{(c)}_{f} -A^{(t)}_{f}  \right)
%\;.
\label{eq:amp}
\end{eqnarray}
For $f=\pi K$, the amplitudes ${\cal A}^{(u)}_{\pi K}$, ${\cal
A}^{(c)}_{\pi K}$, and ${\cal A}^{(t)}_{\pi K}$ are decomposed at
LO into
\begin{eqnarray}
{\cal A}^{(u)}_{\pi K} &=&
f_KF_e+{\cal M}_e+f_\pi F_{eK}+{\cal M}_{eK} + f_BF_a+{\cal M}_a \;,
\nonumber\\
{\cal A}^{(c)}_{\pi K} &=& 0 \;,
\nonumber\\
{\cal A}^{(t)}_{\pi K} &=&
- \left( f_KF^P_e+{\cal M}^P_e +f_\pi F^P_{eK}
        +{\cal M}^P_{eK} +f_BF^P_a+{\cal M}^P_a \right) \;,
\end{eqnarray}
where $f_B$ ($f_K$, $f_\pi$) is the $B$ meson (kaon, pion) decay
constant, $F_e$ (${\cal M}_{e}$) the color-allowed factorizable
(nonfactorizable) tree emission contribution, $F_{eK}$ (${\cal
M}_{eK}$) the color-suppressed factorizable (nonfactorizable) tree
emission contribution, $F_a$ (${\cal M}_{a}$) the factorizable
(nonfactorizable) tree annihilation contribution, and those with
the additional superscripts $P$ the contributions from the penguin
operators. For $f=\pi\pi$, the amplitudes ${\cal
A}^{(u)}_{\pi\pi}$, ${\cal A}^{(c)}_{\pi\pi}$, and ${\cal
A}^{(t)}_{\pi\pi}$ are decomposed at LO into
\begin{eqnarray}
{\cal A}^{(u)}_{\pi\pi} &=&
f_\pi F_e + {\cal M}_e + f_BF_a + {\cal M}_a
\;,
\nonumber\\
{\cal A}^{(c)}_{\pi\pi} &=& 0
\;,
\nonumber\\
{\cal A}^{(t)}_{\pi\pi} &=&
- \left( f_\pi F^P_e + {\cal M}^P_e + f_BF^P_a + {\cal M}^P_a \right)
\;,
\end{eqnarray}
where we do not distinguish the color-allowed and color-suppressed
contributions.

The factorization formulas for the various contributions to each
$B\to \pi K$, $\pi\pi$ mode are collected in Tables~\ref{amp} and
\ref{ampp}, and in Appendix, whose dependence on the Wilson
coefficients has been made explicit. We define the standard
combinations,
\begin{eqnarray}
a_1(\mu) &=& C_2(\mu) + \frac{C_1(\mu)}{N_c}\;,
\nonumber\\
a_2(\mu) &=& C_1(\mu) + \frac{C_2(\mu)}{N_c}\;,
\nonumber\\
a_i(\mu) &=& C_i(\mu) + \frac{C_{i\pm 1}(\mu)}{N_c}\;,
\;\;\;\;i=3\sim  10\;,
\label{com}
\end{eqnarray}
where the upper (lower) sign applies, when $i$ is odd (even). The
coefficients $a$ and $a^{\prime}$ in Tables~\ref{amp} and
\ref{ampp}, besides $a_1$ and $a_2$ given above, are then written
as
\begin{eqnarray}
{\renewcommand\arraystretch{2.2}
\begin{array}{ll}
\displaystyle
a_1'\, =\,  C_1  \;,
&
\displaystyle
a_2'\, =\, C_2 \;,
\\
\displaystyle
a_3^{(q')}\, =\, a_3 + \frac{3}{2} e_{q'} a_9\;,
&
\displaystyle
a_3^{\prime (q')}\, =\, C_4 + \frac{3}{2} e_{q'} C_{10} \;,
\\
\displaystyle
a_4^{(q')}\, =\, a_4 + \frac{3}{2} e_{q'} a_{10} \;,
\ \ \ \
&
\displaystyle
a_4^{\prime (q')}\, =\, C_3 + \frac{3}{2} e_{q'} C_9 \;,
\\
\displaystyle
a_5^{(q')}\, =\, a_5 + \frac{3}{2} e_{q'}  a_7  \;,
&
\displaystyle
a_5^{\prime (q')}\ = \ C_6 + \frac{3}{2} e_{q'} C_8 \;,
\\
\displaystyle
a_6^{(q')}\, =\, a_6 + \frac{3}{2} e_{q'} a_8 \;,
&
\displaystyle
a_6^{\prime (q')}\  =\  C_5 + \frac{3}{2} e_{q'} C_7
\;.
\end{array}
}
\label{wil}
\end{eqnarray}

%%%%%%%%%%%%%%%%%%%%%   Table I  %%%%%%%%%%%%%%%%%%%%%%%%%%%%%%%%
\begin{table}[hb]
\begin{center}
\begin{tabular}{l|cc}
\hline\hline
& ${\cal A}_{\pi^+K^0}^{(u)} $ & $\sqrt{2}{\cal A}_{\pi^0K^+}^{(u)}$
\\\hline
$F_e$ & 0 & $F_{e4} \left( a_1  \right)$
\\
${\cal M}_{e}$ & 0 & ${\cal M}_{e4} \left(  a_1' \right)$
\\
$F_{eK}$ & 0 & $F_{eK4} \left( a_2 \right)$
\\
${\cal M}_{eK}$ & 0 & ${\cal M}_{eK4}\left( a_2' \right)$
\\
$F_a$ & $F_{a4} \left( a_1 \right)$ & $F_{a4} \left( a_1 \right)$
\\
${\cal M}_{a}$ & ${\cal M}_{a4} \left( a_1'\right)$ &
${\cal M}_{a4} \left( a_1'\right)$
\\\hline
& ${\cal A}_{\pi^+K^0}^{(t)} $ & $\sqrt{2}{\cal A}_{\pi^0K^+}^{(t)}$
\\\hline
$F_e^P$ &
$F_{e4} \left( a_4^{(d)} \right) + F_{e6} \left( a_6^{(d)} \right)$ &
$F_{e4} \left( a_4^{(u)} \right) + F_{e6} \left( a_6^{(u)}  \right)$
\\
${\cal M}_{e}^P$ &
${\cal M}_{e4}\left(  a_4^{\prime (d)} \right)
  + {\cal M}_{e6}\left( a_6^{\prime (d)} \right)$ &
${\cal M}_{e4}\left(  a_4^{\prime (u)} \right)
  + {\cal M}_{e6}\left( a_6^{\prime (u)} \right)$
\\
$F_{eK}^P$ & 0 &
$F_{eK4} \left( a_3^{(u)} - a_3^{(d)} - a_5^{(u)} + a_5^{(d)} \right)$
\\
${\cal M}_{eK}^P$ & 0 &
${\cal M}_{eK4}\left( a_3^{\prime (u)} - a_3^{\prime (d)} 
%\right)
%  + {\cal M}_{eK5}\left( 
+ a_5^{\prime (u)} - a_5^{\prime (d)}  \right)$
\\
$F_a^P$ &
$F_{a4} \left( a_4^{(u)} \right) + F_{a6} \left( a_6^{(u)} \right)$ &
$F_{a4} \left( a_4^{(u)} \right) + F_{a6} \left( a_6^{(u)} \right)$
\\
${\cal M}_{a}^P$ &
${\cal M}_{a4}\left( a_4^{\prime (u)} \right)
  + {\cal M}_{a6}\left( a_6 ^{\prime (u)} \right)$ &
${\cal M}_{a4}\left( a_4^{\prime (u)} \right)
  + {\cal M}_{a6}\left( a_6 ^{\prime (u)} \right)$
\\
\hline  & ${\cal A}_{\pi^-K^+}^{(u)}$&
$\sqrt{2}{\cal A}_{\pi^0K^0}^{(u)}$
\\\hline
$F_e$ & $F_{e4} \left( a_1  \right)$ & 0
\\
${\cal M}_{e}$ & ${\cal M}_{e4} \left(  a_1' \right)$ & 0
\\
$F_{eK}$ & 0 & $F_{eK4} \left( a_2 \right)$
\\
${\cal M}_{eK}$ & 0 & ${\cal M}_{eK4}\left( a_2' \right)$
\\
$F_a$ & 0 & 0
\\
${\cal M}_{a}$ & 0 & 0
\\\hline
& ${\cal A}_{\pi^-K^+}^{(t)}$& $\sqrt{2}{\cal A}_{\pi^0K^0}^{(t)}$
\\\hline
$F_e^P$ &
$F_{e4} \left( a_4^{(u)} \right) + F_{e6} \left( a_6^{(u)} \right)$ &
$F_{e4} \left( - a_4^{(d)} \right) + F_{e6} \left( - a_6^{(d)} \right)$
\\
${\cal M}_{e}^P$ &
${\cal M}_{e4}\left(  a_4^{\prime (u)} \right)
  + {\cal M}_{e6}\left( a_6^{\prime (u)} \right)$ &
${\cal M}_{e4}\left(  - a_4^{\prime (d)} \right)
   + {\cal M}_{e6}\left( - a_6^{\prime (d)}  \right)$
\\
$F_{eK}^P$ & 0 &
$F_{eK4} \left( a_3^{(u)} - a_3^{(d)} - a_5^{(u)} + a_5^{(d)} \right)$
\\
${\cal M}_{eK}^P$ & 0 &
${\cal M}_{eK4}\left( a_3^{\prime (u)} - a_3^{\prime (d)} 
%\right)
%   + {\cal M}_{eK5}\left( 
+a_5^{\prime (u)} - a_5^{\prime (d)}  \right)$
\\
$F_a^P$ &
$F_{a4} \left( a_4^{(d)} \right) + F_{a6} \left( a_6^{(d)} \right)$ &
$F_{a4} \left( - a_4^{(d)} \right) + F_{a6} \left( - a_6^{(d)} \right)$
\\
${\cal M}_{a}^P$ &
${\cal M}_{a4}\left( a_4^{\prime (d)} \right)
   + {\cal M}_{a6}\left( a_6 ^{\prime (d)} \right)$ &
${\cal M}_{a4}\left( - a_4^{\prime (d)} \right)
   + {\cal M}_{a6}\left( - a_6 ^{\prime (d)} \right)$
\\\hline\hline
\end{tabular}
\caption{$B\to \pi K$ decay amplitudes, whose factorization
formulas are presented in Appendix.}
\label{amp}
\end{center}
\end{table}
%%%%%%%%%%%%%%%%%%%%%%%%%%%%%%%%%%%%%%%%%%%%%%%%%%%%%%%%%%%%%%%%%

%%%%%%%%%%%%%%%%%%%%%   Table II %%%%%%%%%%%%%%%%%%%%%%%%%%%%%%%%
\begin{table}[hb]
\begin{center}
\begin{tabular}{l|c}
\hline \hline& ${\cal A}_{\pi^+\pi^-}^{(u)} $
\\\hline
$F_e$ & $F_{e4} \left( a_1 \right)$
\\
${\cal M}_{e}$ & ${\cal M}_{e4} \left( a_1' \right)$
\\
$F_a$ & 0
\\
${\cal M}_{a}$ & ${\cal M}_{a4} \left( a_2' \right)$
\\\hline
& ${\cal A}_{\pi^+\pi^-}^{(t)} $
\\\hline
$F_e^P$ &
$F_{e4} \left( a_4^{(u)} \right)+ F_{e6} \left( a_6^{(u)} \right)$
\\
${\cal M}_{e}^P$ & ${\cal M}_{e4}\left( a_4^{\prime (u)} \right)$
\\
$F_a^P$ & $F_{a6} \left(a_6^{(d)} \right)$
\\
${\cal M}_{a}^P$ &
${\cal M}_{a4}
  \left(  a_3^{\prime (u)} + a_3^{\prime (d)} +  a_4^{\prime (d)}
%   \right)
%  + {\cal M}_{a5}\left( 
+ a_5^{\prime (u)} + a_5^{\prime (d)}\right)$
\\\hline   & $\sqrt{2}{\cal A}_{\pi^+\pi^0}^{(u)} $
\\\hline
$F_e$ & $F_{e4} \left( a_1 + a_2 \right)$
\\
${\cal M}_{e}$ & ${\cal M}_{e4} \left( a_1' + a_2' \right)$
\\
$F_a$ & 0
\\
${\cal M}_{a}$ & 0
\\\hline
& $\sqrt{2}{\cal A}_{\pi^+\pi^0}^{(t)} $
\\\hline
$F_e^P$ &
$F_{e4}
  \left( a_3^{(u)} - a_3^{(d)} + a_4^{(u)} - a_4^{(d)}
         - a_5^{(u)} + a_5^{(d)} \right)
  + F_{e6} \left( a_6^{(u)}-a_6^{(d)} \right)$
\\
${\cal M}_{e}^P$ &
${\cal M}_{e4}\left( a_3^{\prime (u)} - a_3^{\prime (d)}
                       + a_4^{\prime (u)} - a_4^{\prime (d)}
% \right)
%  + {\cal M}_{e5} \left( 
+ a_5^{\prime (u)} - a_5^{\prime (d)} \right)$
\\
$F_a^P$   & 0
\\
${\cal M}_{a}^P$   & 0
\\\hline & $\sqrt{2}{\cal A}_{\pi^0\pi^0}^{(u)} $
\\\hline
$F_e$ & $F_{e4} \left(- a_2 \right)$
\\
${\cal M}_{e}$ & ${\cal M}_{e4} \left( -a_2' \right)$
\\
$F_a$ & 0
\\
${\cal M}_{a}$ & ${\cal M}_{a4} \left( a_2' \right)$
\\\hline
& $\sqrt{2}{\cal A}_{\pi^0\pi^0}^{(t)} $
\\\hline
$F_e^P$ &
$ F_{e4} \left( -a_3^{(u)} + a_3^{(d)} +  a_4^{(d)}
                + a_5^{(u)} - a_5^{(d)} \right)
  + F_{e6} \left( a_6^{(d)} \right)$
\\
${\cal M}_{e}^P$ &
${\cal M}_{e4}\left( - a_3^{\prime (u)} + a_3^{\prime (d)}
                     + a_4^{\prime (d)} 
%\right)
%  + {\cal M}_{e5} \left( 
-a_5^{\prime (u)} + a_5^{\prime (d)} \right)$
\\
$F_a^P$ & $F_{a6} \left( a_6^{(d)} \right)$
\\
${\cal M}_{a}^P$ &
${\cal M}_{a4} \left( a_3^{\prime (u)} + a_3^{\prime (d)}
                       +  a_4^{\prime (d)} 
%\right)
%   + {\cal M}_{a5}
%          \left( 
+ a_5^{\prime (u)} + a_5^{\prime (d)}\right)$
\\\hline\hline
\end{tabular}
\caption{$B\to \pi\pi$ decay amplitudes, whose factorization
formulas are presented in Appendix.} \label{ampp}
\end{center}
\end{table}
%%%%%%%%%%%%%%%%%%%%%%%%%%%%%%%%%%%%%%%%%%%%%%%%%%%%%%%%%%%%%%%%%

With the amplitude ${\cal A}(B\to f)$ being computed using
Eq.~(\ref{eq:amp}), we derive the two-body nonleptonic $B$ meson
decay rates and CP asymmetries. The former are given by
\begin{eqnarray}
\Gamma(B\to f)\, =\, \frac{G_F^2m_B^3}{128\pi}|{\cal A}(B\to
f)|^2\;,\label{dra}
\end{eqnarray}
where $m_B$ is the $B$ meson mass. The time-dependent CP asymmetry
of the $B^0\to \pi^0K_S$ mode is defined as
\begin{eqnarray}
A_{CP}(B^0(t)\to \pi^0K_S)
&\equiv&
\frac{B({\bar B}^0(t)\to \pi^0K_S)- B(B^0(t)\to \pi^0K_S)}
{B({\bar B}^0(t)\to \pi^0K_S)+B(B^0(t)\to \pi^0K_S)}
\nonumber\\
&=&
A_{\pi^0K_S}\cos(\Delta M_d\, t)+S_{\pi^0K_S}\sin(\Delta M_d\, t)\;,
\label{CPk}
\end{eqnarray}
with the mass difference $\Delta M_d$ of the two $B$-meson mass
eigenstates, and the direct asymmetry and the mixing-induced
asymmetry,
\begin{eqnarray}
A_{\pi^0K_S}\, =\, {|\lambda_{\pi^0K_S}|^2-1 \over
1+|\lambda_{\pi^0K_S}|^2}\;, \hspace{20mm} S_{\pi^0K_S}\, =\,
{2\,\rm{Im}(\lambda_{\pi^0K_S}) \over
1+|\lambda_{\pi^0K_S}|^2}\;,\label{spk}
\end{eqnarray}
respectively. The $B^0\to\pi^0 K_S$ decay has a CP-odd final
state, and the corresponding factor,
\begin{eqnarray}
\lambda_{\pi^0 K_S}\, =\,
-e^{-2i\phi_1} {P' -P'_{ew} -C'e^{-i\phi_3}
\over P' -P'_{ew} -C'e^{i\phi_3}} \;,\label{mix}
\end{eqnarray}
where the weak phase $\phi_1$ is defined via
$V_{td}=|V_{td}|\exp(-i\phi_1)$. The time-dependent CP asymmetry
of the $B^0\to \pi^+\pi^-$ mode is defined by
\begin{eqnarray}
A_{CP}(B^0(t)\to\pi^+\pi^-)&\equiv&
\frac{B({\bar B}^0(t)\to\pi^+\pi^-)- B(B^0(t)\to\pi^+\pi^-)}
{B({\bar B}^0(t)\to\pi^+\pi^-) +B(B^0(t)\to\pi^+\pi^-)}
\nonumber\\
&=&A_{\pi\pi}\cos(\Delta M_d\, t)+S_{\pi\pi}\sin(\Delta M_d\, t)\;,
\label{CPASY}
\end{eqnarray}
with the direct asymmetry and the mixing-induced asymmetry,
\begin{eqnarray}
A_{\pi\pi}\, =\,
{|\lambda_{\pi\pi}|^2-1 \over 1+|\lambda_{\pi\pi}|^2}\;,
\hspace{20mm}
S_{\pi\pi}\, =\,
{2\,\rm{Im}(\lambda_{\pi\pi}) \over 1+|\lambda_{\pi\pi}|^2}\;,
\end{eqnarray}
respectively, and the factor,
\begin{eqnarray}
\lambda_{\pi\pi}\, =\,
e^{2i\phi_2} {T+Pe^{-i\phi_2} \over T+Pe^{i\phi_2}} \;,\label{mix2}
\end{eqnarray}
where the weak phase $\phi_2$ comes from the identity
$\phi_2=180^\circ-\phi_1-\phi_3$. In addition, the direct CP
asymmetry for a charged $B$ meson decay $B^+\to f(B^-\to \bar f)$
is defined by
\begin{eqnarray}
A_{CP}\, =\,
\frac{B(B^-\to \bar f)- B(B^+\to f)}
{B(B^-\to \bar f) +B(B^+\to f)}
\;.
\end{eqnarray}

The PQCD predictions for the branching ratios and the CP
asymmetries of the $B\to\pi K$, $\pi\pi$ decays in the NDR scheme
are listed in Tables~\ref{br2}-\ref{mixcp}. Using the LO and NLO
Wilson coefficients, we obtain the values in the columns labelled
by LO and LO$_{\rm NLOWC}$, respectively. It is noticed that some
of the NLO Wilson coefficients, like $C_5$, diverge at a low
scale. To derive the above tables, we have frozen the values
$C_i(\mu)$ at $C_i(\mu_0=0.5)$ GeV, whenever $\mu$ runs to below
the scale $\mu_0$, since the renormalization-group (RG) evolution
is not reliable for $\mu <\mu_0$. Note that $\mu_0$ is also the
scale, which sets the starting point of the RG evolution of the
Gegenbauer coefficients in the meson distribution amplitudes
\cite{CZ}. We have kept the corrections in higher orders of the
electroweak coupling $\alpha$ to the Wilson evolution, which were
neglected in \cite{BN}. Because the effect of the NLO Wilson
coefficients is to enhance the penguin amplitude, the branching
ratios of the penguin-dominated $B\to\pi K$ modes increase, and
the magnitudes of the direct CP asymmetries decrease a bit
accordingly. As shown in Eq.~(\ref{Mbpi1}), The enhanced penguin
amplitude $P$, being destructive to the color-allowed tree
amplitude $T$, renders the $B\to\pi\pi$ branching ratios vary
toward the direction favored by the data. The larger subdominant
penguin amplitude also increases the magnitudes of the direct CP
asymmetries in the $B\to\pi\pi$ decays due to the stronger
interference with the dominant tree amplitudes.

As stated before, the LO PQCD predictions for the $B\to\pi K$
branching ratios are consistent with the data, viewing the range
spanned by the columns LO and LO$_{\rm NLOWC}$ in Table~\ref{br2}.
However, the prediction for the $B^0\to\pi^0 \pi^0$ branching
ratio is too small compared to the measured value. Those for the
direct CP asymmetries of the $B\to\pi K$, $\pi\pi$ decays, except
$A_{CP}(B^\pm\to\pi^0 K^\pm)$, are all in good agreement with the
data as shown in Table~\ref{cp2}. The LO direct CP asymmetry of
the $B^0\to\pi^0 \pi^0$ mode differs in sign from the result
obtained in \cite{LUY}, because we have employed the different
pion distribution amplitudes (see Sec.~\ref{TU}). It simply
implies that the theoretical uncertainty for the modes with tiny
branching ratios is huge. Note that the predictions from QCDF
\cite{BBNS} for the direct CP asymmetries usually have signs
opposite to those from PQCD. It has been realized that the set
``S4'' with non-universal parameters, such as the different
annihilation phases for the $B\to PP$, $PV$, and $VP$ decays, must
be adopted in order for QCDF to accommodate the data
\cite{BN,NCKM05,Du,Alek}. The above two discrepancies associated
with the $B^0\to\pi^0 \pi^0$ branching ratio and with the
$B^\pm\to\pi^0 K^\pm$ direct CP asymmetry lead to the puzzles
mentioned in Introduction. We prepare Table~\ref{tp} for the
various topological amplitudes, whose definitions are referred to
\cite{Charng}. The values in the columns LO and LO$_{\rm NLOWC}$
follow the power counting rules in Eqs.~(\ref{po}) and (\ref{pow})
exactly, explaining why the $B\to\pi K$, $\pi\pi$ puzzles appear.

%%%%%%%%%%%%%%%%%%%%%   Table III  %%%%%%%%%%%%%%%%%%%%%%%%%%%%%%
\begin{table}[hbt]
\begin{center}
\begin{tabular}{cccccccc}
\hline\hline Mode & Data \cite{HFAG}& LO & LO$_{\rm NLOWC}$ &
+VC & +QL &  +MP  & +NLO
\\
\hline
$B^\pm \to \pi^\pm K^0$ & $ 24.1 \pm 1.3 $ &
 $17.0$&$32.3$&$31.0$&$34.2$&$24.1$&
 $24.5^{+13.6\,(+12.9)}_{-\ 8.1\,(-\ 7.8)}$
% $17.3$&$32.9$&$31.6$&$34.9$&$24.5$&
% $24.9^{+13.9\,(+13.2)}_{-\ 8.2\,(-\ 8.2)}$
\\
$B^\pm \to \pi^0 K^\pm$ & $ 12.1 \pm 0.8 $ &
 $10.2$&$18.4$&$17.4$&$19.4$&$14.0$&
 $13.9^{+10.0\,(+\ 7.0)}_{-\ 5.6\,(-\ 4.2)}$
% $10.4$&$18.7$&$17.7$&$19.7$&$14.2$&
% $14.2^{+10.2\,(+\ 7.1)}_{-\ 5.8\,(-\ 4.3)}$
\\
$B^0 \to \pi^\mp K^\pm$ & $ 18.9 \pm 0.7 $ &
 $14.2$&$27.7$&$26.7$&$29.4$&$20.5$&
 $20.9^{+15.6\,(+11.0)}_{-\ 8.3\,(-\ 6.5)}$
% $14.3$&$28.0$&$26.9$&$29.7$&$20.7$&
% $21.1^{+15.7\,(+11.1)}_{-\ 8.4\,(-\ 6.6)}$
\\
$B^0 \to \pi^0 K^0 $ & $ 11.5 \pm 1.0 $ &
 $\phantom{0}5.7$&$12.1$&$11.8$&$12.8$&$\phantom{0}8.7$&
 $\phantom{0}9.1^{+\ 5.6\,(+\ 5.1)}_{-\ 3.3\,(-\ 2.9)}$
% $\phantom{0}5.7$&$12.2$&$11.9$&$13.0$& $\phantom{0}8.8$&
% $\phantom{0}9.2^{+\ 5.6\,(+\ 5.1)}_{-\ 3.3\,(-\ 3.0)}$
\\
\hline $B^0 \to \pi^\mp \pi^\pm$ & $ \phantom{0}5.0 \pm 0.4 $ &
 $\phantom{0}7.0$&$\phantom{0}6.8$&$\phantom{0}6.6$&
 $\phantom{0}6.9$&$\phantom{0}6.7$&
 $\phantom{0}6.5^{+\ 6.7\,(+\ 2.7)}_{-\ 3.8\,(-\ 1.8)}$
% $\phantom{0}7.1$& $\phantom{0}6.8$& $\phantom{0}6.6$&
% $\phantom{0}6.9$& $\phantom{0}6.7$&
% $\phantom{0}6.6^{+\ 6.7\,(+\ 2.7)}_{-\ 3.8\,(-\ 1.8)}$
\\
$B^\pm \to \pi^\pm \pi^0$ & $ \phantom{0}5.5 \pm 0.6 $ &
 $\phantom{0}3.5$&$\phantom{0}4.1$&$\phantom{0}4.0$&
 $\phantom{0}4.1$&$\phantom{0}4.1$&
 $\phantom{0}4.0^{+\ 3.4\,(+\ 1.7)}_{-\ 1.9\,(-\ 1.2)}$
% $\phantom{0}3.5$& $\phantom{0}4.2$& $\phantom{0}4.1$&
% $\phantom{0}4.2$& $\phantom{0}4.2$&
% $\phantom{0}4.1^{+\ 3.5\,(+\ 1.7)}_{-\ 2.0\,(-\ 1.2)}$
\\
$B^0 \to \pi^0 \pi^0$ & $ \phantom{0}1.45 \pm 0.29 $ &
 $\phantom{0}0.12$&$\phantom{0}0.27$&$\phantom{0}0.37$&
 $\phantom{0}0.29$&$\phantom{0}0.21$&
 $\phantom{0}0.29^{+0.50\,(+0.13)}_{-0.20\,(-0.08)}$
% $\phantom{0}0.12$&$\phantom{0}0.28$&$\phantom{0}0.37$&
% $\phantom{0}0.29$&$\phantom{0}0.21$&
% $\phantom{0}0.30^{+0.49\,(+0.12)}_{-0.21\,(-0.09)}$
\\
\hline\hline
\end{tabular}
\end{center}
\caption{Branching ratios in the NDR scheme ($\times 10^{-6}$).
The label LO$_{\rm NLOWC}$ means the LO results with the NLO
Wilson coefficients, and +VC, +QL, +MP, and +NLO mean the
inclusions of the vertex corrections, of the quark loops, of the
magnetic penguin, and of all the above NLO corrections,
respectively. The errors in the parentheses were defined in the
context.}\label{br2}
\end{table}
%%%%%%%%%%%%%%%%%%%%%%%%%%%%%%%%%%%%%%%%%%%%%%%%%%%%%%%%%%%%%%%%%

%%%%%%%%%%%%%%%%%%%%%   Table IV  %%%%%%%%%%%%%%%%%%%%%%%%%%%%%%%
\begin{table}[hbt]
\begin{center}
\begin{tabular}{cccccccl}
\hline\hline Mode & Data \cite{HFAG}& LO & LO$_{\rm NLOWC}$&
+VC & +QL &  +MP  & \ \ \ \ \ \ \ \ +NLO
\\
\hline
$B^\pm \to \pi^\pm K^0$ & $ -0.02 \pm  0.04$ &
 $-0.01$&$-0.01$&$-0.01$& $\phantom{-}0.00$&$-0.01$&
 $\phantom{-}0.00\pm 0.00\,(\pm 0.00)$
\\
$B^\pm \to \pi^0 K^\pm$ & $ \phantom{-}0.04 \pm 0.04 $ &
 $-0.08$&$-0.06$&$-0.01$&$-0.05$&$-0.08$&
 $-0.01^{+0.03\,(+0.03)}_{-0.05\,(-0.05)}$
\\
$B^0 \to \pi^\mp K^\pm$ & $ -0.115 \pm 0.018 $ &
 $-0.12$&$-0.08$&$-0.09$&$-0.06$&$-0.10$&
 $-0.09^{+0.06\,(+0.04)}_{-0.08\,(-0.06)}$
\\
$B^0 \to \pi^0 K^0 $ & $ \phantom{-}0.02 \pm 0.13 $ &
%--- &
 $-0.02$& $\phantom{-}0.00$&$-0.07$& $\phantom{-}0.00$&
 $\phantom{-}0.00$&
 $-0.07^{+0.03\,(+0.01)}_{-0.03\,(-0.01)}$
\\
\hline $B^0 \to \pi^\mp \pi^\pm$ & $ \phantom{-}0.37 \pm 0.10$ &
 $\phantom{-}0.14$& $\phantom{-}0.19$& $\phantom{-}0.21$&
 $\phantom{-}0.16$& $\phantom{-}0.20$&
 $\phantom{-}0.18^{+0.20\,(+0.07)}_{-0.12\,(-0.06)}$
\\
$B^\pm \to \pi^\pm \pi^0$ & $ \phantom{-}0.01 \pm 0.06 $&
 $\phantom{-}0.00$& $\phantom{-}0.00$& $\phantom{-}0.00$&
 $\phantom{-}0.00$& $\phantom{-}0.00$&
 $\phantom{-}0.00\pm 0.00\,(\pm 0.00)$
\\
$B^0 \to \pi^0 \pi^0$ & $ \phantom{-}0.28^{+0.40}_{-0.39}$ &
 $-0.04$&$-0.34$& $\phantom{-}0.65$&$-0.41$&$-0.43$&
 $\phantom{-}0.63^{+0.35\,(+0.09)}_{-0.34\,(-0.15)}$
\\
\hline\hline
\end{tabular}
\end{center}
\caption{Direct CP asymmetries in the NDR scheme.}\label{cp2}
\end{table}
%%%%%%%%%%%%%%%%%%%%%%%%%%%%%%%%%%%%%%%%%%%%%%%%%%%%%%%%%%%%%%%%%

%%%%%%%%%%%%%%%%%%%%%   Table VI  %%%%%%%%%%%%%%%%%%%%%%%%%%%%%%%
\begin{table}[hbt]
\begin{center}
\begin{tabular}{cllllll}
\hline\hline Topology &\ \ \ LO & LO$_{\rm NLOWC}$ &
\ \ +VC &\ \ +QL &\ \  +MP &\ +NLO
\\
\hline
$P'$  &
  $36.6\, e^{i\,2.9} $ & $50.6\, e^{i\,2.9} $ & $49.6\, e^{i\,2.9} $ &
  $52.1\, e^{i\,2.9} $ & $43.7\, e^{i\,2.8} $ & $44.1\, e^{i\,2.9} $
\\
$T'$ &
  $\phantom{0}6.9\, e^{i\,0.0} $ & $\phantom{0}6.6\, e^{i\,0.0} $ &
  $\phantom{0}6.6\, e^{i\,0.1} $ & $\phantom{0}6.6\, e^{i\,0.0} $ &
  $\phantom{0}6.6\, e^{i\,0.0} $ & $\phantom{0}6.6\, e^{i\,0.1} $
\\
$C'$ &
  $\phantom{0}0.5\, e^{-i\,2.5} $ & $\phantom{0}0.6\, e^{-i\,0.4} $ &
  $\phantom{0}1.9\, e^{-i\,1.3} $ & $\phantom{0}0.6\, e^{-i\,0.2} $ &
  $\phantom{0}0.6\, e^{-i\,0.4} $ & $\phantom{0}1.7\, e^{-i\,1.3} $
\\
$P'_{ew}$  &
  $\phantom{0}5.8\, e^{i\,3.1} $ & $\phantom{0}5.8\, e^{-i\,3.1} $ &
  $\phantom{0}5.4\, e^{-i\,3.0} $ & $\phantom{0}5.8\, e^{-i\,3.1} $ &
  $\phantom{0}5.8\, e^{-i\,3.1} $ & $\phantom{0}5.4\, e^{-i\,3.0} $
\\
\hline
$T$  &
  $24.3\, e^{i\,0.0} $ & $23.5\, e^{i\,0.0} $ & $23.1\, e^{i\,0.0} $ &
  $23.6\, e^{-i\,0.1} $ & $23.5\, e^{i\,0.0}$ & $23.2\, e^{i\,0.0} $
\\
$P$ &
  $\phantom{0}4.7\, e^{-i\,0.4} $ & $\phantom{0}6.5\, e^{-i\,0.4} $ &
  $\phantom{0}6.3\, e^{-i\,0.3} $ & $\phantom{0}6.7\, e^{-i\,0.3} $ &
%  $\phantom{0}5.7\, e^{-i\,0.4} $ & $\phantom{0}5.6\, e^{-i\,0.3} $
  $\phantom{0}5.7\, e^{-i\,0.4} $ & $\phantom{0}5.6\, e^{-i\,0.4} $
\\
$C$ &
  $\phantom{0}0.8\, e^{i\,2.6} $ & $\phantom{0}2.2\, e^{i\,0.2} $ &
  $\phantom{0}4.8\, e^{-i\,1.1} $ & $\phantom{0}2.3\, e^{i\,0.4} $ &
  $\phantom{0}2.2\, e^{i\,0.2} $ & $\phantom{0}4.3\, e^{-i\,1.1} $
\\
$P_{ew}$ &
  $\phantom{0}0.7\, e^{i\,0.0} $ & $\phantom{0}0.7\, e^{i\,0.0} $ &
  $\phantom{0}0.7\, e^{-i\,0.1} $ & $\phantom{0}0.7\, e^{i\,0.0} $ &
  $\phantom{0}0.7\, e^{i\,0.0} $ & $\phantom{0}0.7\, e^{-i\,0.1} $
\\
\hline\hline
\end{tabular}
\end{center}
\caption{Topological amplitudes in units of $10^{-5}$ GeV for the
$B\to\pi K$, $\pi\pi$ decays in the NDR scheme.}\label{tp}
\end{table}
%%%%%%%%%%%%%%%%%%%%%%%%%%%%%%%%%%%%%%%%%%%%%%%%%%%%%%%%%%%%%%%%%

%%%%%%%%%%%%%%%%%%%%%   Table V  %%%%%%%%%%%%%%%%%%%%%%%%%%%%%%%%
\begin{table}[hbt]
\begin{center}
\begin{tabular}{cccccccc}
\hline\hline  & Data & LO & LO$_{\rm NLOWC}$& +VC & +QL &  +MP  &
+NLO
\\
\hline $S_{\pi^0 K_S}$ &
 $\phantom{-}0.31\pm0.26$&
% $ 0.34^{+0.27}_{-0.29}$&
 $\phantom{-}0.70$& $\phantom{-}0.73$& $\phantom{-}0.74$&
 $\phantom{-}0.73$& $\phantom{-}0.73$&
 $\phantom{-}0.74^{+0.02\,(+0.01)}_{-0.03\,(-0.01)}$
\\
$S_{\pi\pi}$ &$ - 0.50 \pm 0.12 $ &
 $-0.34$&$-0.49$&$-0.47$&$-0.51$&$-0.41$&
 $-0.42^{+1.00\,(+0.05)}_{-0.56\,(-0.05)}$
\\
\hline\hline
\end{tabular}
\end{center}
\caption{Mixing-induced CP asymmetries in the NDR scheme.
%with $\phi_1=21.6^\circ$.
}\label{mixcp}
\end{table}
%%%%%%%%%%%%%%%%%%%%%%%%%%%%%%%%%%%%%%%%%%%%%%%%%%%%%%%%%%%%%%%%%

After obtaining the values of the various topological amplitudes,
we compute the mixing-induced CP asymmetries through
Eqs.~(\ref{mix}) and (\ref{mix2}). Since $C'$ is of $O(\lambda^2)$
compared to $P'$, it is expected that the LO PQCD results of
$S_{\pi^0K_S}$ are close to that extracted from the $b\to c\bar
cs$ decays, $S_{c\bar cs}=\sin(2\phi_1)\approx 0.685$, as shown in
Table~\ref{mixcp}. On the contrary, $P$ is of $O(\lambda)$ of $T$
in the $B^0\to\pi^\mp\pi^\pm$ decays, such that a larger deviation
of $S_{\pi\pi}$ from $S_{c\bar cs}$ is found. The LO PQCD results
of $S_{\pi\pi}$ are consistent with the data, but those of
$S_{\pi^0K_S}$ are not. Moreover, PQCD predicts $\Delta S_{\pi^0
K_S}\equiv S_{\pi^0K_S}-S_{c\bar cs}> 0$, opposite to the observed
values. This result is in agreement with those obtained in the
literature \cite{CGRS,CCS2,B05}. Hence, the measurement of the
mixing-induced CP asymmetries in the penguin-dominated modes
provides an opportunity of discovering new physics. Currently, the
data of $S_{\pi^0K_S}$ still suffer significant errors. On the
other hand, the NLO corrections and the theoretical uncertainty,
which concern the allowed range of the PQCD predictions, need to
be analyzed. A more clear picture will be attained, after we
complete these analyses.

\section{NEXT-TO-LEADING-ORDER CORRECTIONS}

We explain the consistent power countings in $\alpha_s$ and in
large logarithms, before computing the NLO corrections. A PQCD
formula of leading-power in $1/m_b$ is written symbolically as
\begin{eqnarray}
  \exp[\gamma^{(0)}(\alpha_s) L_{WC}]\otimes
  \exp[\Gamma^{(0)}(\alpha_s)L_S^2+\Gamma^{(1)}(\alpha_s^2)L_S^2]
  \otimes \exp[\gamma_q^{(0)}(\alpha_s) L_{RG}]\otimes
  H^{(0)}(\alpha_s)\otimes\phi(\mu_0)\;,\label{1}
\end{eqnarray}
where the first, second, and third exponentials represent the
Wilson coefficient, the Sudakov factor, and the RG factor
\cite{CL}, with the notations $L_{WC}\equiv \ln(m_W/t)$,
$L_S\equiv \ln(xP b)$, and $L_{RG}\equiv\ln(tb)$, $xP$ being a
fractional parton momentum, $t\sim\sqrt{\bar\Lambda m_b}$ a
characteristic hard scale, and $b$ the conjugate variable to the
parton transverse momentum $k_T$, and $\gamma$, $\Gamma$, and
$\gamma_q$ the corresponding anomalous dimensions. The RG factor
governs the evolution from $t$ down to $1/b$. The evolution from
$1/b$ down to the cutoff $\mu_0$, which characterizes the meson
distribution amplitude $\phi$, has been neglected. This formula is
complete at LO, since the hard kernel $H$ is evaluated to
$O(\alpha_s)$, and at next-to-leading logarithm (NLL), since the
Wilson coefficient, the Sudakov factor, and the RG factor have
been resummed up to the next-to-leading logarithms
$\alpha_sL_{WC}$, $\alpha_s L_S$,  and $\alpha_s L_{RG}$,
respectively ($\alpha_s L_S^2$ is the leading logarithm). In all
our previous works we used the one-loop running coupling constant
$\alpha_s$, which is, strictly speaking, a NLO effect. This effect
takes into account the potential large NLO corrections to the $B$
meson transition form factors through the BLM procedure (see
Introduction).

Next, we add subleading corrections to Eq.~(\ref{1}), which
include
\begin{enumerate}
\item $H^{(0)}(\alpha_s) \to
H^{(0)}(\alpha_s)+H^{(1)}(\alpha_s^2)$. This is what we are going
to do in this section, where the NLO hard kernel $H^{(1)}$
contains the vertex corrections, the quark loops, and the magnetic
penguin.

\item  $\exp[\gamma^{(0)}(\alpha_s) L_{WC}] \to
\exp[\gamma^{(0)}(\alpha_s) L_{WC}+\gamma^{(1)}(\alpha_s^2)
L_{WC}]$. The LO Wilson coefficient is replaced by the NLO one,
for which the corresponding anomalous dimension is calculated to
two loops: $\gamma^{(0)}(\alpha_s) \to \gamma^{(0)}(\alpha_s)
+\gamma^{(1)}(\alpha_s^2)$. According to our counting rules, the
NLO anomalous dimension leads to the summation of the
next-to-next-to-leading logarithm (NNLL) $\alpha_s^2L_{WC}$.

\item
$\exp[\Gamma^{(0)}(\alpha_s)L_S^2+\Gamma^{(1)}(\alpha_s^2)L_S^2]
\to \exp[\Gamma^{(0)}(\alpha_s)L_S^2+\Gamma^{(1)}(\alpha_s^2)L_S^2
+\Gamma^{(2)}(\alpha_s^3)L_S^2]$. This means the accuracy of the
summation up to NNLL ($\alpha_s^3L_S^2$). Unfortunately, it
requires a three-loop evaluation of the corresponding anomalous
dimension for the Sudakov factor, which is not yet available in
the literature.

\item $\exp[\gamma_q^{(0)}(\alpha_s) L_{RG}] \to
\exp[\gamma_q^{(0)}(\alpha_s) L_{RG}
+\gamma_q^{(1)}(\alpha_s^2)L_{RG}]$. Since $L_{RG}$ and $L_S$ are
of the same order of magnitude, and the NNLL Sudakov resummation
is not available, this NNLL piece of subleading corrections
($\alpha_s^2L_{RG}$) can not be included consistently.

\end{enumerate}

The power countings in $\alpha_s$ and in various large logarithms
are independent in principle. Based on the above classification,
we shall extend Eq.~(\ref{1}) by considering the subleading
corrections from the first and second pieces. With the one-loop
running $\alpha_s$, the NLO corrections to the hard kernel are
complete (assuming that the corrections to the form factors have
been minimized by our choice of the hard scale). It is not
necessary to adopt the two-loop $\alpha_s$ as in \cite{BN}, whose
effect is next-to-next-to-leading-order (NNLO). Because of $L_{WC}
\gg L_S, L_{RG}$, the NNLL term $\alpha_s^2 L_{WC}$ is much more
essential than those from the third and fourth pieces. The LO PQCD
results for the $B\to\pi K$, $\pi\pi$ decays from using the NLO
Wilson coefficients have been listed in
Tables~\ref{br2}-\ref{mixcp}. When investigating the NLO
corrections from the vertex corrections, the quark loops, and the
magnetic penguin to the hard kernel below, we shall always use the
NLO Wilson coefficients. After obtaining the decay amplitudes
${\cal A}(B\to f)$ up to NLO, we employ Eq.~(\ref{dra}) to
evaluate the corresponding decay rates.

\subsection{Vertex Corrections}

It has been known that the vertex corrections, reducing the
dependence of the Wilson coefficients on the renormalization scale
$\mu$, play an important role in a NLO analysis. Since the
nonfactorizable contributions are negligible \cite{KL}, we
concentrate only on the vertex corrections to the factorizable
amplitudes. For charmless $B$ meson decays, these corrections do
not involve the end-point singularities from vanishing momentum
fractions in collinear factorization theorem (QCDF \cite{BBNS}).
Therefore, there is no need to employ $k_T$ factorization theorem
(PQCD \cite{KLS,LUY,CL,LY1,YL}) here. This claim can be justified
by recalculating one of the nonfactorizable amplitudes, ${\cal
M}_{e4}$, for the $B\to \pi K$ decays in collinear factorization
theorem, which is also free of the end-point singularity. It is
found that the results for ${\cal M}_{e4}$ from the two
calculations (with and without the parton transverse momentum
$k_T$ in the kaon) differ only by 10\%. For more detail, refer to
Appendix. After justifying the neglect of the parton transverse
degrees of freedom, we simply quote the QCDF expressions for the
vertex corrections. An important remark is that the light quark
from the $b$ quark transition is assumed to carry the full
momentum of the associated meson in QCDF \cite{BBNS}. Strictly
speaking, this light quark carries the fractional momentum, whose
dependence should appear in the PQCD formalism for the vertex
corrections. Because it is indeed an energetic quark, the
assumption is reasonable.

The vertex corrections modify the Wilson coefficients in
Eq.~(\ref{com}) into \cite{BBNS}
\begin{eqnarray}
a_1(\mu) &\to& a_1(\mu)
+\frac{\alpha_s(\mu)}{4\pi}C_F\frac{C_{1}(\mu)}{N_c} V_1(M) \;,
\nonumber\\
a_2(\mu) &\to& a_2(\mu)
+\frac{\alpha_s(\mu)}{4\pi}C_F\frac{C_{2}(\mu)}{N_c} V_2(M) \;,
\nonumber\\
a_i(\mu) &\to& a_i(\mu)
+\frac{\alpha_s(\mu)}{4\pi}C_F\frac{C_{i\pm 1}(\mu)}{N_c} V_i(M)
\;,\;\;\;\;i=3 - 10,
\end{eqnarray}
with $M$ being the meson emitted from the weak vertex. For the
$B\to\pi K$ decays, $M$ denotes the kaon for the vertex functions
$V_{1,4,6,8,10}$ and the pion for $V_{2,3,5,7,9}$. In the NDR
scheme $V_i(M)$ are given by \cite{BBNS}
\begin{eqnarray}
V_i(M) &=& \left\{
{\renewcommand\arraystretch{2.5}
\begin{array}{ll}
12\ln\displaystyle{\frac{m_b}{\mu}}-18
+\frac{2\sqrt{2N_c}}{f_M}\int_0^1 dx\, \phi_M^A(x)\, g(x)\;, &
\mbox{\rm for }i=1-4,9,10\;,
\\
-12\ln\displaystyle{\frac{m_b}{\mu}}+6
-\frac{2\sqrt{2N_c}}{f_M}\int_0^1dx\, \phi_M^A(x)\, g(1-x)\;, &
\mbox{\rm for }i=5,7\;,
\\
\displaystyle{ -6 +\frac{2\sqrt{2N_c}}{f_M}\int_0^1 dx\,
\phi_M^{P}(x)\, h(x) }\;, & \mbox{\rm for }i=6,8\;,
\end{array}
}
\right.\label{vim}
\end{eqnarray}
where $f_M$ is the decay constant of the meson $M$, and
$\phi_M^A(x)$ [$\phi_M^P(x)$] the twist-2 (twist-3) meson
distribution amplitude given in Sec.~\ref{TU}, $x$ being the
parton momentum fraction. The hard kernels are
\begin{eqnarray}
g(x) &=& 3\left( \frac{1-2x}{1-x}\ln{x} -i\,\pi \right)\nonumber\\
& & +\left[ 2\,{\rm Li}_2(x)-\ln^2 x +\frac{2\ln
x}{1-x}-(3+2i\,\pi)\ln x - (x\leftrightarrow 1-x) \right] \;,
\\
h(x) &=& 2\,{\rm Li}_2(x)-\ln^2 x -(1+2i\,\pi)\ln x -
(x\leftrightarrow 1-x) \;.
\end{eqnarray}
The expressions of $V_i(M)$ in the HV scheme can be found in
\cite{Cheng}. The factorization formulas for the various $B\to\pi
K$, $\pi\pi$ decay amplitudes are still the same as in
Tables~\ref{amp} and \ref{ampp}.

\begin{figure}[tb]
\begin{center}
\begin{tabular}{ccc}
 \includegraphics[scale=0.8]{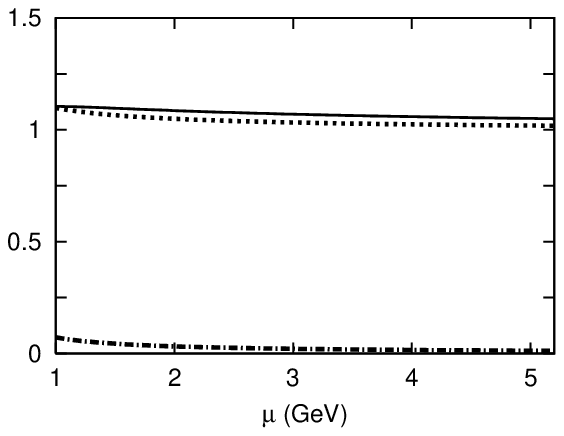}
&
 \includegraphics[scale=0.8]{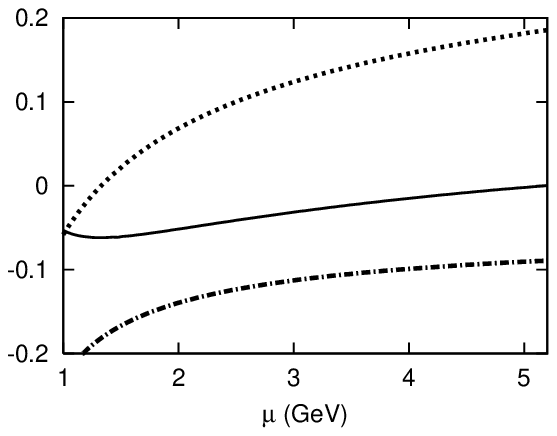}
&
 \includegraphics[scale=0.8]{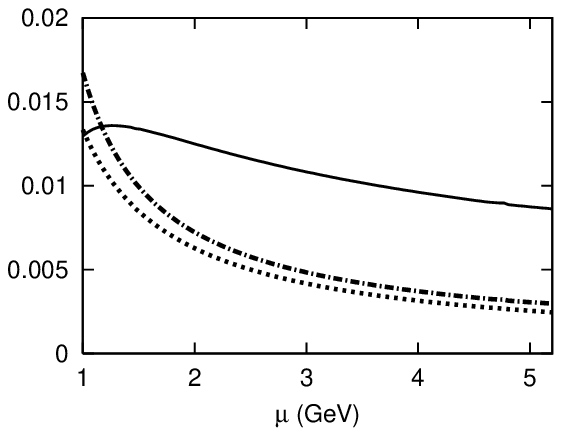}
\\
(a) $a_1$
&
(b) $a_2$
&
(c) $a_3$
\\
 \includegraphics[scale=0.8]{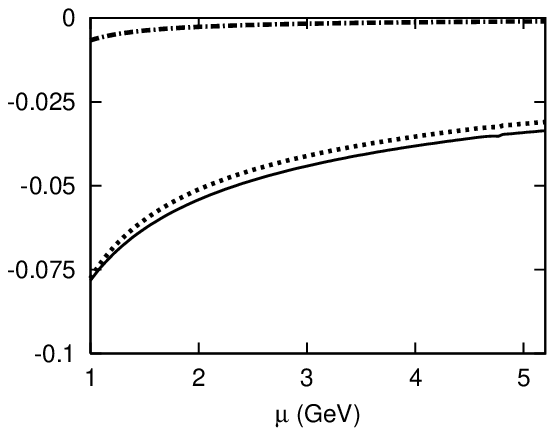}
&
 \includegraphics[scale=0.8]{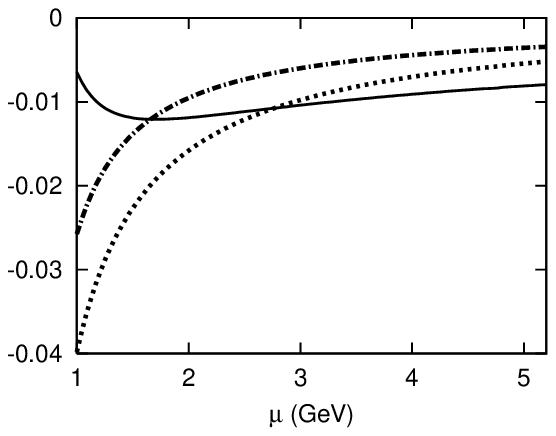}
&
 \includegraphics[scale=0.8]{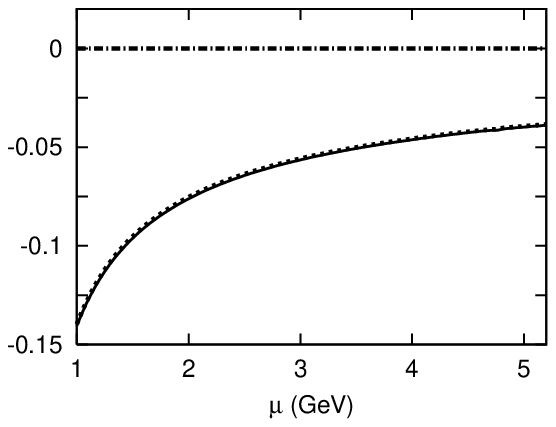}
\\
(d) $a_4$
&
(e) $a_5$
&
(f) $a_6$
\\
 \includegraphics[scale=0.8]{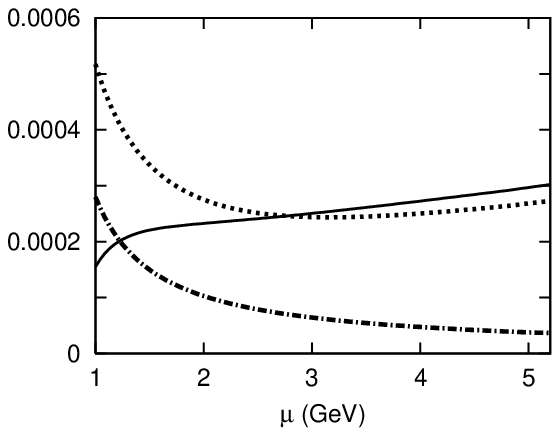}
&
 \includegraphics[scale=0.8]{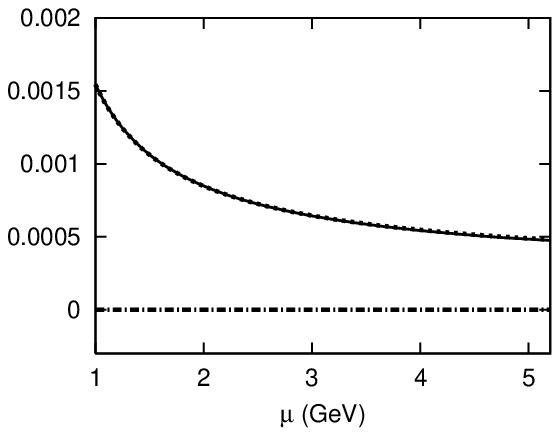}
&
 \includegraphics[scale=0.8]{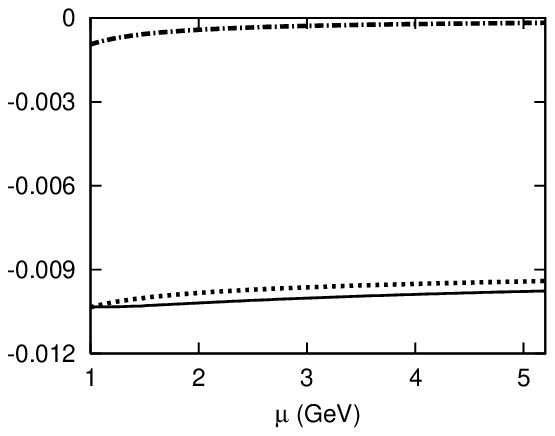}
\\
(g) $a_7$
&
(h) $a_8$
&
(i) $a_9$
\\
 \includegraphics[scale=0.8]{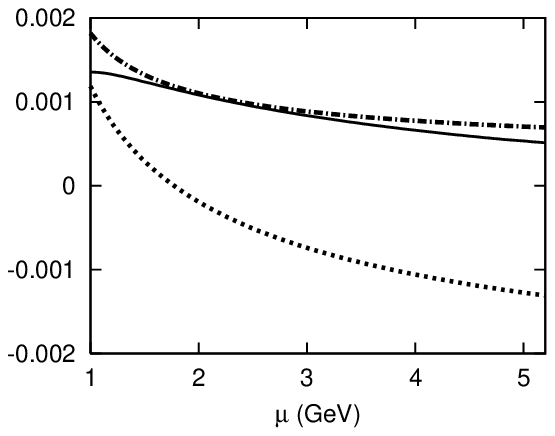}
\\
(g) $a_{10}$
\end{tabular}
\caption{Real parts of $a_i$ for the $B\to \pi K$ decays without
the vertex corrections (dotted lines) and with the vertex
corrections (solid lines), and imaginary parts with the vertex
corrections (dot-dashed lines) in the NDR scheme. }\label{fig:vc}
\end{center}
\end{figure}

The dependence of the Wilson coefficients $a_i(\mu)$ on the
renormalization scale $\mu$ modified by the vertex corrections is
exhibited in Fig.~\ref{fig:vc} for both the real and imaginary
parts. It is found that the $\mu$ dependence of most of $a_i$ is
moderated by the vertex corrections (with the generation of the
imaginary parts). The $\mu$ dependence of $a_{6,8}$ is, however,
not altered. It has been known that their dependence will be
moderated after combined with the $\mu$ dependence of the chiral
scale $m_{0K}(\mu)$ associated with the kaon \cite{BBNS}. The most
dramatic changes arise from $a_{2,3,10}$. Due to the smallness of
$a_3$ ($a_{10}$) compared to the Wilson coefficient $a_{4,6}$
($a_9$) for the QCD (electroweak) penguins, the only significant
effect appears in the color-suppressed tree amplitude $C'$, which
is governed by $a_2$. For other $a_i$, the vertex corrections
amount only up to 70\% at the scale $\mu\sim \sqrt{\bar\Lambda
m_b}\sim 1.5$ GeV. The above observation is manifest in
Table~\ref{tp}: most of the topological amplitudes for the
$B\to\pi K$, $\pi\pi$ decays change a little, while $C'$ and $C$
are enhanced by factors of 3 and 2 (viewing the values in the
columns LO$_{\rm NLOWC}$ and +VC ), respectively.

It is then understood that the $B\to\pi K$ branching ratios,
dominated by the penguin contributions from $a_{4,6}$, vary
only slightly under the vertex corrections, as indicated in
Table~\ref{br2}.
However, the direct CP asymmetries of the $B^\pm\to\pi^0 K^\pm$
and $B^0 \to \pi^0 K^0$ modes, related to $C'$, are modified
significantly, as shown in Tables~\ref{cp2}: % and \ref{cp1}:
$A_{CP}(B^\pm\to\pi^0 K^\pm)$ has increased from $-0.06$ to
$-0.01$, and $A_{CP}(B^0 \to \pi^0 K^0)\equiv A_{\pi^0K_S}$ has
decreased from $0.00$ to $-0.07$. $A_{CP}(B^0\to \pi^\mp K^\pm)$,
determined solely by the color-allowed tree amplitude $T'$, does
not change much. The effect from the vertex corrections on the LO
PQCD predictions for the $B\to\pi\pi$ decays can also be
understood by means of the enhanced color-suppressed tree
amplitude $C$: the $B^0\to \pi^0\pi^0$ branching ratio increases
by 30\%, and the direct CP asymmetry changes from $-0.34$ to
$+0.65$. The sign flip of the direct CP asymmetry is attributed to
a huge change of the strong phase of $C$ caused by the vertex
corrections. The predicted $B^0\to\pi^\pm\pi^\mp$ and
$B^\pm\to\pi^\pm\pi^0$ branching ratios, to which $C$ remains
subdominant, decrease only a bit. The NLO effect, though
increasing $|C|$ by a factor of 2, is not enough to resolve the
$B\to\pi\pi$ puzzle. Perhaps, the penguin amplitude is also larger
than expected \cite{BPRS,ALP}. Nevertheless, the vertex
corrections do improve the consistency between the theoretical
predictions and the experimental data of the $B\to\pi\pi$ decays.

%%%%%%%%%%%%%%%%%%%%%   Figure 1  %%%%%%%%%%%%%%%%%%%%%%%%%%%%%%%
%%%%%%%%%%%%%%%%%%%%%%%%%%%%%%%%%%%%%%%%%%%%%%%%%%%%%%%%%%%%%%%%%

Though the vertex corrections have been included in QCDF
\cite{BBNS}, they do not help resolve the $B\to \pi K$ puzzle. We
neglect the electroweak penguin $P'_{ew}$ for convenience in the
following explanation. Table~\ref{tp} shows that the penguin
amplitude $P'$ is in the second quadrant, and the color-allowed
tree amplitude $T'$ is roughly aligned with the positive real
axis. The color-suppressed tree amplitude $C'$ is enhanced by the
vertex corrections, and becomes almost imaginary. It then orients
the sum $T'+C'$ into the fourth quadrant, such that $T'+C'$ and
$P'$ more or less line up (and point to the opposite directions).
This is the reason the magnitude of $A_{CP}(B^\pm \to \pi^0
K^\pm)$, proportional to the sine of the angle between $T'+C'$ and
$P'$, becomes smaller in PQCD. The situation in QCDF is different,
where $P'$ is preferred to be in the third quadrant \cite{KL}.
That is, the predicted $A_{CP}(B^0 \to \pi^\mp K^\pm)$ has a wrong
sign. Then the modified $C'$, still orienting $T'+C'$ into the
fourth quadrant, can not reduce the magnitude of $A_{CP}(B^\pm \to
\pi^0 K^\pm)$. The three types of NLO corrections considered here
have been extended up to $O(\alpha_s^2\beta_0)$ in QCDF recently
\cite{BW0504}, which, however, make the QCDF predictions for
$A_{CP}(B^\pm \to \pi^0 K^\pm)$ more deviate from the data.
Another $O(\alpha_s^2)$ piece from the $b\to s g^*g^*$ transition
was included into QCDF \cite{LY0508}, which enhances the $B\to\pi
K$ branching ratios, but leaves their direct CP asymmetries
intact. The $B\to\pi K$ puzzle can not be resolved in SCET either
\cite{BPS05}: the leading SCET formalism requires the ratio
$C'/T'$ to be real, so that $C'$, being parallel to $T'$, can not
orient the sum $T'+C'$ into the fourth quadrant, and that the
magnitude of $A_{CP}(B^\pm \to \pi^0 K^\pm)$ remains large.

We have found that the color-suppressed tree amplitude $C'$ could
be enhanced few times by the vertex corrections in the standard
model. It is then worthwhile to investigate whether the
mixing-induced CP asymmetry $S_{\pi^0K_S}$ in the $B\to \pi^0K_S$
decays deviates from $S_{c\bar cs}$ substantially under a large
$C'$ according to Eq.~(\ref{mix}). A similar investigation of the
large $C$ effect applies to $S_{\pi\pi}$ in the $B^0 \to \pi^\mp
\pi^\pm$ decays according to Eq.~(\ref{mix2}). The results are
collected in Table~\ref{mixcp}, which indicates that the deviation
 is still small and positive. It is known
that the leading deviation caused by $C'$ is proportional to
$\cos(\delta_{C'}-\delta_{P'})$, if neglecting $P'_{ew}$. Because
the vertex corrections also rotate the orientation of $C'$, it
becomes more orthogonal to $P'$ as shown in Table~\ref{tp}, and
$\Delta S_{\pi^0 K_S}$ is not increased much. The mixing-induced
CP asymmetry $S_{\pi\pi}$, depending only on $T$ and $P$, is
insensitive to the vertex corrections, which mainly affect $C$.

\subsection{Quark Loops}

For the $B\to \pi K$ and $B\to\pi \pi$ decays, the dominant
penguin amplitude $P'\sim |V_{tb}V^*_{ts}|C_4$ and tree amplitude
$T\sim |V_{ub}V^*_{ud}|C_2$ are both of $O(\lambda^4)$
\cite{Charng2}. Hence, the charm-quark loop amplitude,
proportional to $\alpha_s|V_{cb}V^*_{cs}|C_2\sim
\alpha_s\lambda^2$ in the former and to
$\alpha_s|V_{cb}V^*_{cd}|C_2\sim \alpha_s\lambda^3$ in the latter,
could be an important source of NLO corrections. Its effect is
expected to be larger in the $B\to \pi K$ decays. On the other
hand, the up-quark loop amplitude, proportional to
$\alpha_s|V_{ub}V^*_{us}|C_2\sim \alpha_s\lambda^5$ \cite{Charng2}
for $B\to \pi K$, seems to be negligible. For $B\to \pi \pi$, the
up-quark loop amplitude, proportional to
$\alpha_s|V_{ub}V^*_{ud}|C_2\sim \alpha_s\lambda^4$
\cite{Charng2}, might be comparable to the charm-quark one.
Therefore, we shall include both quark loops in the following
analysis. For completeness, we shall also include the quark-loop
amplitudes from the QCD penguin operators, whose contributions are
proportional to $\alpha_s|V_{tb}V^*_{ts}|C_i\sim
\alpha_s\lambda^4$, $i=3,4,6$. They have the order of magnitude
the same as or larger than the up-quark one, and can influence the
direct CP asymmetries of the $B\to\pi K$ modes. The quark loops
from the electroweak penguins will be neglected due to their
smallness. Note that the CKM factors of these corrections differ
among the loop amplitudes and between the $b\to s(d)$ and $\bar
b\to \bar s(\bar d)$ transitions.

\begin{figure}[tb]
\begin{center}
\begin{tabular}{cc}
\includegraphics{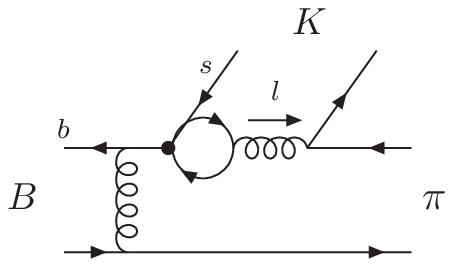} \hspace{10mm}
& \includegraphics{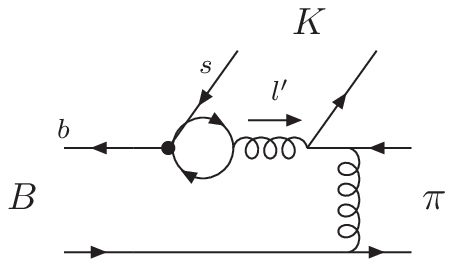}
\\
(a)\hspace{10mm} & (b)
\end{tabular}
\caption{Quark-loop amplitudes.}\label{loops}
\end{center}
\end{figure}

For the $b\to s$ transition, the contributions from the various
quark loops are given by
\begin{eqnarray}
H_{\rm eff}\, =\, - \sum_{q=u,c,t}\sum_{q'}
\frac{G_F}{\sqrt{2}}V_{qb}V^*_{qs}
 \frac{\alpha_s(\mu)}{2\pi}
C^{(q)}(\mu,l^2) \left( \bar s
    \gamma_\rho
    (1-\gamma_5) T^a b
\right) \left( \bar q'\gamma^\rho T^a q' \right)\;,
\label{eq:cp}
\end{eqnarray}
$l^2$ being the invariant mass of the gluon, which attaches the
quark loops in Fig.~\ref{loops}. For the $b\to d$ transition, the
quark-loop corrections are obtained by substituting $d$ for $s$ in
Eq.~(\ref{eq:cp}). The functions $C^{(q)}(\mu,l^2)$ are written as
\begin{eqnarray}
C^{(q)}(\mu,l^2)\, =\, \left[ G^{(q)}(\mu,l^2)  - \frac{2}{3}%\kappa
\right] C_2(\mu) \;,
\end{eqnarray}
for $q=u$, $c$, and
\begin{eqnarray}
C^{(t)}(\mu,l^2)\, =\, \left[ G^{(s)}(\mu,l^2) - \frac{2}{3}%\kappa
\right] C_3(\mu)\ + \sum_{q''=u,d,s,c} G^{(q'')}(\mu,l^2) \left[
C_4(\mu) + C_6(\mu) \right] \;.\label{ctt}
\end{eqnarray}
%with the parameter $\kappa=1$ ($\kappa=0$) in the NDR (HV) scheme.
The constant term $-2/3$ in the above expressions arises from the
Fierz transformation of the four-fermion operators in $D$
dimensions with the anti-commuting Dirac matrix $\gamma_5$ in the
NDR scheme. The contribution proportional to the Wilson
coefficient $C_5$, being purely ultraviolet, should be combined
with that from the magnetic penguin to form the effective Wilson
coefficient $C_{8g}+C_{5}$ \cite{REVIEW}. Since our characteristic
hard scale is of order $\sqrt{\bar\Lambda m_b}\sim 1.5$ GeV, the
$b$ quark is not an active one, and does not contribute to
Eq.~(\ref{ctt}). Except this difference, our expressions are
basically the same as in \cite{BBNS}.

%%%%%%%%%%%%%%%%%%%%%   Figure 2  %%%%%%%%%%%%%%%%%%%%%%%%%%%%%%%
%%%%%%%%%%%%%%%%%%%%%%%%%%%%%%%%%%%%%%%%%%%%%%%%%%%%%%%%%%%%%%%%%

The function $G^{(c)}(\mu,l^2)$ for the loop of the massive charm
quark is given by
\begin{eqnarray}
G^{(c)}(\mu,l^2) &=& - 4 \int_0^1 dx\, x (1-x) \ln\frac{m_c^2 -
x(1-x) l^2}{\mu^2} \;,\label{gc}
\end{eqnarray}
$m_c$ being the charm quark mass, whose real and imaginary parts
are
\begin{eqnarray}
{\rm Re}\, G^{(c)}(\mu, l^2) &=& \frac{2}{3} \left( \frac{5}{3} +
\frac{4m_c^2}{l^2} -\ln\frac{m_c^2}{\mu^2} \right)\nonumber\\
& & + \frac{2}{3} \left(
  1+\frac{2m_c^2}{l^2}
\right) \left\{
\begin{array}{ll}
\sqrt{1-\frac{4m_c^2}{l^2}}
\ln\frac{\sqrt{1-\frac{4m_c^2}{l^2}}-1}{\sqrt{1-\frac{4m_c^2}{l^2}}+1}\;,
& -\infty < l^2 < 0
\\
- 2\sqrt{\frac{4m_c^2}{l^2}-1} \cot^{-1} \left(
  \sqrt{\frac{4m_c^2}{l^2}-1}
\right)\;, & 0 \leq l^2 < 4 m_c^2
\\
-2 \left(1-\frac{4m_c^2}{l^2} \right)\;, & l^2 = 4 m_c^2
\\
\sqrt{1-\frac{4m_c^2}{l^2}}
\ln\frac{1-\sqrt{1-\frac{4m_c^2}{l^2}}}{1+\sqrt{1-\frac{4m_c^2}{l^2}}}\;,
& 4 m_c^2 < l^2 < \infty
\end{array}
\right.
\end{eqnarray}
and
\begin{eqnarray}
{\rm Im}\, G^{(c)}(\mu,l^2) &=& \frac{2\pi}{3} \left(
  1+\frac{2m_c^2}{l^2}
\right) \sqrt{1-\frac{4m_c^2}{l^2}}\, \theta \left(
  1-\frac{4m_c^2}{l^2}
\right) \;,
\end{eqnarray}
respectively. For the loops of the light quarks $u, d$, and $s$,
we have the expressions similar to Eq.~(\ref{gc}) but with $m_c$
being replaced by $m_u$, $m_d$, and $m_s$, respectively.
%\begin{eqnarray}
%G^{(u,d,s)}(\mu,l^2) &=& - 4 \int_0^1 dx\, x (1-x) \ln\frac{- x(1-x)
%(l^2 + i\,\epsilon)}{\mu^2}
%\nonumber\\
%&=& \left\{
%{\renewcommand\arraystretch{2.5}
%\begin{array}{ll}
%\displaystyle
%\frac{2}{3} \left[
%\frac{5}{3} -\ln\frac{-l^2}{\mu^2} \right]\;, & l^2 < 0
%\\
%\displaystyle
%\frac{2}{3} \left[ \frac{5}{3} -\ln\frac{l^2}{\mu^2} + i\,\pi
%\right]\;, & l^2 >0\;.
%\end{array}
%}
%\right.
%\end{eqnarray}
%and
%\begin{eqnarray}
%G^{(u,d,s)}(\mu,l^2=0) &=&
%- \lim_{m^2\to 0} \frac{2}{3} \ln\frac{m^2}{\mu^2}\;.
%\end{eqnarray}
Because their contributions are insensitive to the light quark
masses, we simply adopt the same mass $m$ for the three quark
loops. Varying $m$ from $m_u=4.5$ MeV to $m_s\approx 100$ MeV, the
branching ratios change by less than 1\%.

%A small mass $m$ is necessary for facilitating the numerical
%evaluation of the contributions from the light quark loops, such
%as $m=m_u$, where $m_u=0.0045$ GeV is the $u$ quark mass. We have
%made sure that our numerical results are stable with respect to
%the variation of $m$.

\begin{figure}[tb]
\begin{center}
\begin{tabular}{cc}
 \includegraphics[scale=0.9]{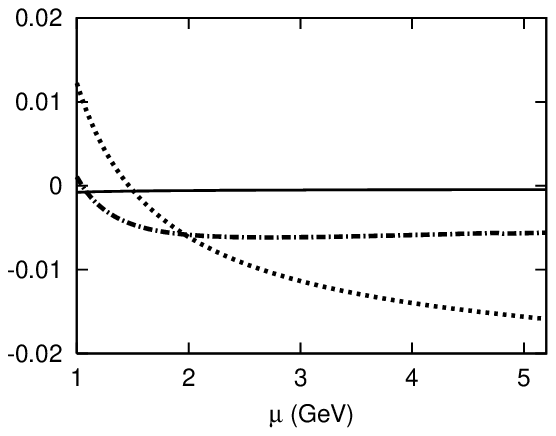}
&
 \includegraphics[scale=0.9]{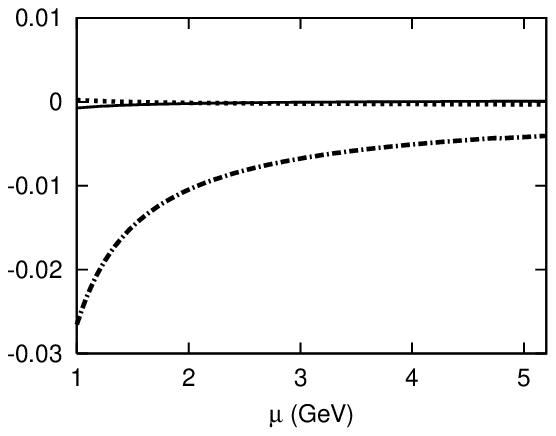}
\\
(a) Re$\left[\frac{\alpha_s(\mu)}{9\pi}\,
\frac{V_{qb}V^*_{qs}}{V_{tb}V^*_{ts}}\,C^{(q)}(\mu,l^2)\right]$ &
(b) Im$\left[\frac{\alpha_s(\mu)}{9\pi}\,
\frac{V_{qb}V^*_{qs}}{V_{tb}V^*_{ts}}\,C^{(q)}(\mu,l^2)\right]$
\\
 \includegraphics[scale=0.9]{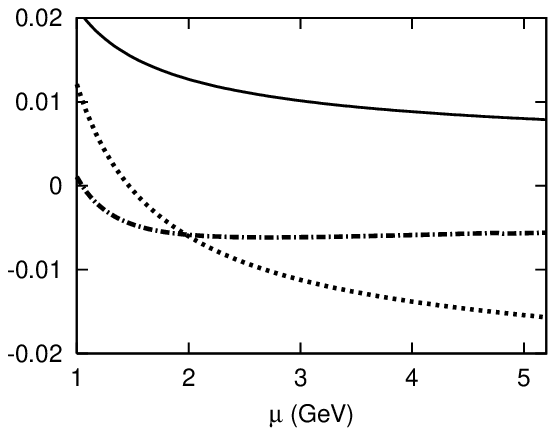}
&
 \includegraphics[scale=0.9]{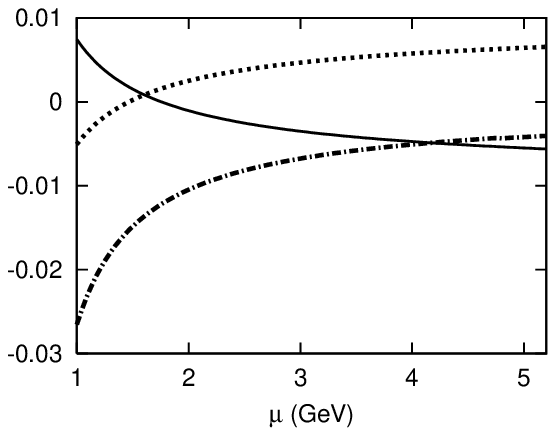}
\\
(c) Re$\left[\frac{\alpha_s(\mu)}{9\pi}\,
\frac{V_{qb}V^*_{qd}}{V_{tb}V^*_{td}}\,C^{(q)}(\mu,l^2)\right]$ &
(d) Im$\left[\frac{\alpha_s(\mu)}{9\pi}\,
\frac{V_{qb}V^*_{qd}}{V_{tb}V^*_{td}}\,C^{(q)}(\mu,l^2)\right]$
\\
\end{tabular}
\caption{Quark-loop contributions to the $b\to s$[(a) and (b)] and
$b\to d$[(c) and (d)] transitions for $l^2=m_B^2/4$ with the
solid, dotted, and dot-dashed lines corresponding to the up-quark,
charm-quark, and QCD-penguin loops, respectively.
}\label{fig:loop}
\end{center}
\end{figure}

To picture the quark-loop effect, we display in
Fig.~\ref{fig:loop} the dependence of
$[V_{qb}V^*_{qs(d)}/(V_{tb}V^*_{ts(d)})] C^{(q)}$, $q=u,c,t$, on
the renormalization scale $\mu$ for a given $l^2=m_B^2/4$ in the
NDR scheme. The real part of the up-quark loop is indeed
negligible compared to that of the charm-quark loop in the $b\to
s$ transition as indicated in Fig.~\ref{fig:loop}(a). However, in
the other transitions described by
Figs.~\ref{fig:loop}(b)-\ref{fig:loop}(d), the up- and charm-loop
corrections are comparable as argued above. The quark loops from
the QCD penguin operators are in fact essential.
Figures~\ref{fig:loop}(a) and \ref{fig:loop}(c) [and also
Figs.~\ref{fig:loop}(b) and \ref{fig:loop}(d)] imply that the weak
phases cause different $\mu$ dependences between the $b\to s$ and
$b\to d$ transitions in the cases of the up and charm loops, but
not in the case of the QCD-penguin loops.

%%%%%%%%%%%%%%%%%%%%%   Figure 3  %%%%%%%%%%%%%%%%%%%%%%%%%%%%%%%
%%%%%%%%%%%%%%%%%%%%%%%%%%%%%%%%%%%%%%%%%%%%%%%%%%%%%%%%%%%%%%%%%

The quark-loop amplitudes depend on the gluon invariant mass
$l^2$, which is assumed to be an arbitrary constant $\langle
l^2\rangle$ in FA. Since the topology displayed in
Fig.~\ref{loops} is the same as the penguin one, its contribution
can be absorbed into the Wilson coefficients $a_{4,6}$,
\begin{eqnarray}
a_{4,6}(\mu) &\to& a_{4,6}(\mu) +
\frac{\alpha_s(\mu)}{9\pi}\sum_{q=u,c,t}\frac{V_{qb}V^*_{qs(d)}}
{V_{tb}V^*_{ts(d)}}
C^{(q)}(\mu,\langle l^2\rangle) \;, \label{eq:CPfact}
\end{eqnarray}
with the other $a_i$ unmodified. The resultant values of $a_{4,6}$
at $\mu=1.5$ and 4.4 GeV are listed in Table~\ref{tab:pen2}. As
$\mu=1.5$ GeV, the quark-loop corrections do not change $a_{4,6}$
much for $b\to s$ and $\bar b\to \bar s$, while they are
destructive (constructive) to $a_{4,6}$ for $b\to d$ ($\bar b\to
\bar d$). As $\mu=4.4$ GeV, these corrections are always
constructive for the different $b$ quark transitions. Besides, the
quark-loop corrections are mode-dependent. For example, they
cancel between the $u\bar u$ and $d\bar d$ components of
$\pi^0=(u\bar u-d\bar d)/\sqrt{2}$ in the $B^\pm\to\pi^\pm\pi^0$
decays, but do not in others.

%%%%%%%%%%%%%%%%%%%%%   Table VII  %%%%%%%%%%%%%%%%%%%%%%%%%%%%%%
\begin{table}[hbt]
\begin{center}
\begin{tabular}{l|cccc}
\hline & LO$_{\rm NLOWC}$ & +QL ($b\to s$) & +QL ($b\to d$) & +MP
\\
\hline $a_4$($1.5$ GeV) &
$-0.0601$ &
$-0.0659 -i\,0.0152$ & $-0.0500 -i\,0.0131$ & $-0.0492$
%$-0.0632 -i\,0.0152$ & $-0.0472 -i\,0.0131$ & $-0.0492$
\\
$a_6$($1.5$ GeV) &
$-0.0952$ &
$-0.1010 -i\,0.0152$ & $-0.0850 -i\,0.0131$ & $-0.0843$
%$-0.0982 -i\,0.0152$ & $-0.0823 -i\,0.0131$ & $-0.0843$
\\
\hline
$a_4$($4.4$ GeV) &
$-0.0336$ &
$-0.0545 -i\,0.0048$ & $-0.0454 -i\,0.0036$ & $-0.0279$
%$-0.0546 -i\,0.0048$ & $-0.0455 -i\,0.0036$ & $-0.0279$
\\
$a_6$($4.4$ GeV) &
$-0.0428$ &
$-0.0637 -i\,0.0048$ & $-0.0546 -i\,0.0036$ & $-0.0371$
%$-0.0638 -i\,0.0048$ & $-0.0547 -i\,0.0036$ & $-0.0371$
\\
\hline & LO$_{\rm NLOWC}$ & +QL ($\bar b\to \bar s$) &
+QL ($ \bar b\to \bar d$)& +MP
\\
\hline
$a_4$($1.5$ GeV) &
$-0.0601$ &
$-0.0646 -i\,0.0150$ & $-0.0804 -i\,0.0180$ & $-0.0492$
%$-0.0618 -i\,0.0150$ & $-0.0776 -i\,0.0180$ & $-0.0492$
\\
$a_6$($1.5$ GeV) &
$-0.0952$ &
$-0.0997 -i\,0.0150$ & $-0.1155 -i\,0.0180$ & $-0.0843$
%$-0.0969 -i\,0.0150$ & $-0.1127 -i\,0.0180$ & $-0.0843$
\\
\hline
$a_4$($4.4$ GeV) &
$-0.0336$ &
$-0.0537 -i\,0.0047$ & $-0.0628 -i\,0.0065$ & $-0.0279$
%$-0.0538 -i\,0.0047$ & $-0.0630 -i\,0.0047$ & $-0.0279$
\\
$a_6$($4.4$ GeV) &
$-0.0428$ &
$-0.0630 -i\,0.0047$ & $-0.0720 -i\,0.0065$ & $-0.0371$
%$-0.0628 -i\,0.0065$ & $-0.0720 -i\,0.0065$ & $-0.0371$
\\
\hline
\end{tabular}
\caption{$a_{4,6}$ including the quark loops and the magnetic
penguin for $l^2=m_B^2/4$ in the NDR scheme.} \label{tab:pen2}
\end{center}
\end{table}
%%%%%%%%%%%%%%%%%%%%%%%%%%%%%%%%%%%%%%%%%%%%%%%%%%%%%%%%%%%%%%%%%

The assumption of a constant gluon invariant mass in FA introduces
a large theoretical uncertainty as making predictions. In the more
sophisticated PQCD approach, the gluon mass is related to the
parton momenta unambiguously (see Appendix). Due to the absence
of the end-point singularities associated with $l^2, l^{\prime
2}\to 0$ in Figs.~\ref{loops}(a) and \ref{loops}(b), respectively,
we have dropped the parton transverse momenta $k_T$ in $l^2,
l^{\prime 2}$ for simplicity. The amplitudes in Eq.~(\ref{eq:amp})
become
\begin{eqnarray}
{\renewcommand\arraystretch{2.0}
\begin{array}{ll}
\displaystyle
{\cal A}^{(u,c)}_{\pi^+K^0 } \, \to\, {\cal A}^{(u,c)}_{\pi^+K^0
}+{\cal M}^{(u,c)}_{\pi K}\;,
&\displaystyle
{\cal A}^{(t)}_{\pi^+K^0 }
\, \to\, {\cal A}^{(t)}_{\pi^+K^0 }-{\cal M}^{(t)}_{\pi K}\;,
\\
\displaystyle
{\cal A}^{(u,c)}_{\pi^0K^+ } \, \to\, {\cal A}^{(u,c)}_{\pi^0K^+ }
+\frac{1}{\sqrt{2}}{\cal M}^{(u,c)}_{\pi K}\;,
&\displaystyle
 {\cal A}^{(t)}_{\pi^0K^+ } \, \to\, {\cal A}^{(t)}_{\pi^0K^+ }
-\frac{1}{\sqrt{2}}{\cal M}^{(t)}_{\pi K}\;,
\\
\displaystyle
{\cal A}^{(u,c)}_{\pi^-K^+ } \, \to\, {\cal A}^{(u,c)}_{\pi^-K^+ }
+{\cal M}^{(u,c)}_{\pi K}\;,
& \displaystyle
{\cal A}^{(t)}_{\pi^-K^+ }
\, \to\, {\cal A}^{(t)}_{\pi^-K^+ } -{\cal M}^{(t)}_{\pi K}\;,
\\
\displaystyle
{\cal A}^{(u,c)}_{\pi^0K^0 } \, \to\, {\cal A}^{(u,c)}_{\pi^0K^0 }
-\frac{1}{\sqrt{2}}{\cal M}^{(u,c)}_{\pi K}\;,
& \displaystyle
{\cal A}^{(t)}_{\pi^0K^0 } \, \to\, {\cal A}^{(t)}_{\pi^0K^0 }
+\frac{1}{\sqrt{2}}{\cal M}^{(t)}_{\pi K}\;,
\\
\displaystyle
{\cal A}^{(u,c)}_{\pi^+\pi^-} \, \to\, {\cal A}^{(u,c)}_{\pi^+\pi^-}
+{\cal M}^{(u,c)}_{\pi\pi}\;,
& \displaystyle
{\cal A}^{(t)}_{\pi^+\pi^-} \, \to\, {\cal A}^{(t)}_{\pi^+\pi^-}
 - {\cal M}^{(t)}_{\pi\pi}\;,
\\
\displaystyle
{\cal A}^{(u,c,t)}_{\pi^+\pi^0} \, \to\, {\cal A}^{(u,c,t)}_{\pi^+\pi^0}\;,
&
\\
\displaystyle
{\cal A}^{(u,c)}_{\pi^0\pi^0} \, \to\, {\cal A}^{(u,c)}_{\pi^0\pi^0}
  + \frac{1}{\sqrt{2}}{\cal M}^{(u,c)}_{\pi\pi}\;,
&\displaystyle
 {\cal A}^{(t)}_{\pi^0\pi^0} \, \to\,
  {\cal A}^{(t)}_{\pi^0\pi^0}-\frac{1}{\sqrt{2}}{\cal M}^{(t)}_{\pi\pi}\;,
\end{array}
}
\end{eqnarray}
where ${\cal M}_f^{(u)}$, ${\cal M}_f^{(c)}$ and ${\cal
M}_f^{(t)}$ denote the up-, charm-, and QCD-penguin-loop
corrections, respectively, and the minus sign for the final state
$\pi^0K^0 $ comes from the $d\bar d$ component in $\pi^0$. The
factorization formulas for ${\cal M}^{(u, c, t)}_{\pi K}$ and
${\cal M}^{(u, c, t)}_{\pi \pi}$ are presented in Appendix.

As indicated in Eq.~(\ref{eq:CPfact}), the quark-loop corrections
affect the penguin contributions, but have a minor impact on other
topological amplitudes. This observation is clear in
Table~\ref{tp}: $|P'|$ ($|P|$) has increased from 50.6 to 52.1
(6.5 to 6.7) in the NDR scheme. Since the $B\to\pi K$ decays are
penguin-dominated, their branching ratios receive an enhancement
(see Table~\ref{br2}). The increase of the branching ratios then
reduces the magnitude of the direct CP asymmetries in the $B\to\pi
K$ modes slightly as shown in Table~\ref{cp2}. It is also easy to
understand the insensitivity of the mixing-induced CP asymmetry
$S_{\pi^0K_S}$ to the quark-loop corrections (see
Table~\ref{mixcp}), viewing the small change in the dominant
amplitude $P'$ in Eq.~(\ref{mix}). On the contrary, the penguin
contribution is subdominant in the $B\to\pi\pi$ decays, so the
branching ratios do not vary much. However, the direct CP
asymmetries $A_{CP}(B^0\to\pi^\mp\pi^\pm)$ and $A_{CP}(B^0\to
\pi^0\pi^0)$, and the mixing-induced CP asymmetry $S_{\pi\pi}$,
directly related to the penguin amplitude, change sizably.

\subsection{Magnetic Penguins\label{sec:MC}}

We then discuss the NLO corrections from the magnetic penguin,
whose weak effective Hamiltonian contains the $b\to sg$
transition,
\begin{eqnarray}
H_{\rm eff}\, =\, -\frac{G_F}{\sqrt{2}}
V_{tb}V_{ts}^*C_{8g}O_{8g}\;,\label{8g}
\end{eqnarray}
with the magnetic penguin operator,
\begin{eqnarray}
O_{8g}\, =\, \frac{g}{8\pi^2}m_b{\bar
s}_i\sigma_{\mu\nu}(1+\gamma_5)T_{ij}^aG^{a\mu\nu}b_j\;,
\end{eqnarray}
$i$, $j$ being the color indices. The Hamiltonian for the $b\to d$
transition is obtained by changing $s$ into $d$ in Eq.~(\ref{8g}).
The topology of the magnetic penguin operator is similar to that
of the quark loops. If regarding the invariant mass $l^2$ of the
gluon emitted from the operator $O_{8g}$ as a constant $\langle
l^2\rangle$, the magnetic-penguin contribution to the $B\to\pi K$,
$\pi\pi$ decays can also be included into the Wilson coefficients,
similar to Eq.~(\ref{eq:CPfact}),
\begin{eqnarray}
a_{4,6}(\mu)&\to& a_{4,6}(\mu) - \frac{\alpha_s(\mu)}{9\pi}
\frac{2m_B}{\sqrt{\langle l^2\rangle}}C_{8g}^{\rm eff}(\mu)\;,
\end{eqnarray}
with the effective Wilson coefficient $C_{8g}^{\rm eff}=
C_{8g}+C_5$ \cite{REVIEW}.
The resultant Wilson coefficients $a_{4,6}(\mu)$ for $\mu=1.5$ and
4.4 GeV have been presented in Table~\ref{tab:pen2}. The
cancellation between the real parts of the quark-loop corrections
and of the magnetic penguin is obvious, except in the case of the
$b\to d$ transition for $\mu=1.5$ GeV.

In the PQCD approach the gluon invariant mass $l^2$ is related to
the parton momenta, such that the corresponding factorization
formulas involve the convolutions of all three meson distribution
amplitudes. Because the nonfactorizable contributions are
negligible, we calculate only the magnetic-penguin corrections to
the the factorizable amplitudes, which modify only ${\cal
A}^{(t)}_{f}$ in Eq.~(\ref{eq:amp}):
\begin{eqnarray}
{\cal A}^{(t)}_{\pi^+K^0 } &\to&{\cal A}^{(t)}_{\pi^+K^0 }
- {\cal M}^{(g)}_{\pi K}
\;, \nonumber\\
{\cal A}^{(t)}_{\pi^0K^+ } &\to&{\cal A}^{(t)}_{\pi^0K^+ }
-\frac{1}{\sqrt{2}}{\cal M}^{(g)}_{\pi K}
\;, \nonumber\\
{\cal A}^{(t)}_{\pi^-K^+ } &\to&{\cal A}^{(t)}_{\pi^-K^+ }
-{\cal M}^{(g)}_{\pi K}
\;, \nonumber\\
{\cal A}^{(t)}_{\pi^0K^0 } &\to&{\cal A}^{(t)}_{\pi^0K^0 } +
\frac{1}{\sqrt{2}}{\cal M}^{(g)}_{\pi K}
\;,\nonumber\\
{\cal A}^{(t)}_{\pi^+\pi^-} &\to& {\cal A}^{(t)}_{\pi^+\pi^-}
- {\cal M}^{(g)}_{\pi \pi}
\;, \nonumber\\
{\cal A}^{(t)}_{\pi^+\pi^0} &\to& {\cal A}^{(t)}_{\pi^+\pi^0}
\;, \nonumber\\
{\cal A}^{(t)}_{\pi^0\pi^0} &\to& {\cal A}^{(t)}_{\pi^0\pi^0}-
\frac{1}{\sqrt{2}}{\cal M}^{(g)}_{\pi \pi} \;.
\end{eqnarray}
The explicit expressions for the magnetic-penguin amplitudes
${\cal M}^{(g)}_{\pi K}$ and ${\cal M}^{(g)}_{\pi \pi}$ are
referred to Appendix. Since an end-point singularity arises, as
the invariant mass $l^2$ approaches zero, we have employed the
$k_T$ factorization theorem, i.e., the PQCD approach in this case.

The effect of the magnetic penguin is just opposite to that of the
quark-loop corrections as indicated in Tables~\ref{br2}-\ref{tp}:
it decreases all the $B\to\pi K$, $\pi\pi$ branching ratios,
except those of the tree-dominated $B^0\to\pi^\mp\pi^\pm$ and
$B^+\to\pi^\pm\pi^0$ modes, and intends to increase the magnitude
of most of the direct CP asymmetries. The mixing-induced CP
asymmetry $S_{\pi^0K_S}$ is stable under the magnetic-penguin
correction for the same reason. The magnitude of $S_{\pi\pi}$
decreases due to the smaller penguin pollution. Because the
quark-loop corrections are smaller than the magnetic penguin, the
pattern of their combined effect is similar to that of the latter.
In summary, the above two pieces of NLO corrections reduce the LO
penguin amplitudes by about 10\% in the $B\to\pi K$, $\pi\pi$
decays, and the $B\to\pi K$ and $B^0\to\pi^0\pi^0$ branching
ratios by about 20\%. The direct CP asymmetries are not altered
very much, which are mainly affected by the vertex corrections, as
shown by the similarities between the columns +VC and +NLO in
Table~\ref{cp2}.

\section{THEORETICAL UNCERTAINTY\label{TU}}

In this section we explain in detail how to derive the results in
Tables~\ref{br2}-\ref{mixcp}, and discuss their theoretical
uncertainty. The PQCD predictions depend on the inputs for the
nonperturbative parameters, such as decay constants, distribution
amplitudes, and chiral scales for pseudoscalar mesons. For the $B$
meson, the model wave function has been proposed in \cite{TLS}:
\begin{eqnarray}
\phi_B(x,b)&=&N_Bx^2(1-x)^2
\exp\left[-\frac{1}{2}\left(\frac{xm_B}{\omega_B}\right)^2
-\frac{\omega_B^2 b^2}{2}\right] \;, \label{os}
\end{eqnarray}
where the Gaussian form was motivated by the oscillator model in
\cite{BW}, and the normalization constant $N_B$ is related to the
decay constant $f_B$ through
\begin{eqnarray}
\int_0^1 dx\, \phi_B(x,b=0)&=&\frac{f_B}{2\sqrt{2N_c}}\;.
\end{eqnarray}
There are certainly other models of the $B$ meson wave function
available in the literature (see \cite{LL04,HQW}). It has been
confirmed that the model in Eq.~(\ref{os}) and the model derived
in \cite{KQT} with a different functional form lead to similar
numerical results for the $B\to\pi$ transition form factor
\cite{WY}.

The twist-2 pion (kaon) distribution amplitude $\phi^A_{\pi(K)}$,
and the twist-3 ones $\phi_{\pi(K)}^P$ and $\phi_{\pi(K)}^T$ have
been parameterized as
\begin{eqnarray}
\phi_{\pi(K)}^A(x) &=& \frac{f_{\pi(K)}}{2\sqrt{2N_c}}\, 6x(1-x)
\left[1 + a_1^{\pi(K)} C_1^{3/2}(2x-1) +
a_2^{\pi(K)}C_2^{3/2}(2x-1)+a_4^{\pi(K)}C_4^{3/2}(2x-1)\right] \;,
\\
\phi^P_{\pi(K)}(x) &=& \frac{f_{\pi(K)}}{2\sqrt{2N_c}}\, \bigg[ 1
+\left(30\eta_3 -\frac{5}{2}\rho_{\pi(K)}^2\right) C_2^{1/2}(2x-1)
\nonumber\\
& & \hspace{35mm} -\, 3\left\{ \eta_3\omega_3 +
\frac{9}{20}\rho_{\pi(K)}^2(1+6a_2^{\pi(K)}) \right\}
C_4^{1/2}(2x-1) \bigg]\;,
\\
\phi^T_{\pi(K)}(x) &=& \frac{f_{\pi(K)}}{2\sqrt{2N_c}}\,
(1-2x)\bigg[ 1 + 6\left(5\eta_3 -\frac{1}{2}\eta_3\omega_3 -
\frac{7}{20}
      \rho_{\pi(K)}^2 - \frac{3}{5}\rho_{\pi(K)}^2 a_2^{\pi(K)} \right)
(1-10x+10x^2) \bigg]\;,\ \ \ \
\end{eqnarray}
with $a_1^\pi=0$, the mass ratio
$\rho_{\pi(K)}=(m_u+m_{d(s)})/m_{\pi(K)}=m_{\pi(K)}/m_{0\pi(K)}$
and the Gegenbauer polynomials $C_n^{\nu}(t)$,
\begin{eqnarray}
&\displaystyle C_2^{1/2}(t)\, =\, \frac{1}{2} \left(3\, t^2-1\right)
\;,&
C_4^{1/2}(t)\, =\, \frac{1}{8} \left(3-30\, t^2+35\, t^4\right) \;,
\nonumber\\
&\displaystyle C_1^{3/2}(t)\, =\, 3\, t \;, &
C_2^{3/2}(t)\, =\, \frac{3}{2} \left(5\, t^2-1\right) \;,
\;\;\;\;\;\;
C_4^{3/2}(t) \,=\, \frac{15}{8} \left(1-14\, t^2+21\, t^4\right) \;.
\end{eqnarray}
In the above kaon distribution amplitudes the momentum fraction
$x$ is carried by the $s$ quark. For both the pion and kaon, we
choose $\eta_3=0.015$ and $\omega_3=-3$ \cite{PB2}. Because we did
not employ the equations of motions for the twist-3 meson
distribution amplitudes \cite{BBNS}, we are allowed to include the
higher Gegenbauer terms, which are in fact important. However, we
drop the derivative term with respect to the transverse parton
momentum $k_T$ in $\phi^T_{\pi(K)}$. It has been observed that the
contribution from this derivative term to the $B\to\pi$ form
factor is negligible \cite{HW04}.

%For the kaon distribution amplitudes, the $s$ quark in $K^-$ and
%the $u$ quark in $K^+$ carry the momentum fraction $x$.

In our previous works we adopted the models of the pion and kaon
distribution amplitudes derived from QCD sum rules in \cite{PB2}.
Fixing the $B$ meson decay constant $f_B\approx 190$ MeV from
lattice QCD (see \cite{LRev}), the shape parameter of the $B$
meson wave function was determined to be $\omega_B\approx 0.4$ GeV
\cite{TLS} from the $B\to\pi$ form factor $F_+^{B\pi}(0)\approx
0.3$ in light-cone sum rules \cite{KR,PB3}. The chiral scales were
chosen as $m_{0\pi}\approx 1.3$ GeV for the pion and
$m_{0K}\approx 1.7$ GeV for the kaon \cite{KLS}. The
renormalization scale $\mu$ was set to the off-shellness of the
internal particles, consistent with the BLM procedure. The
resultant PQCD predictions \cite{KLS} have been confirmed by the
observed $B\to\pi K$ branching ratios and $B^0\to\pi^\mp K^\pm$
direct CP asymmetry. The consistency indicates not only that the
above inputs are reasonable, but that the short-distance QCD
dynamics has been described correctly in PQCD.

In this paper we employ the updated models of the pion and kaon
distribution amplitudes in \cite{Ball:2004ye}. Since the updated
Gegenbauer coefficient $a_2^\pi=0.115$ is smaller than the
previous one 0.44 for the twist-2 pion distribution amplitude
\cite{PB2}, $F_+^{B \pi}(0)$ reduces compared to that obtained in
\cite{KS}. To compensate this reduction, we increase the $B$ meson
decay constant up to $f_B=210$ MeV, which is consistent with the
recent lattice result \cite{AG0507}, in order to maintain the
$B\to\pi K$, $\pi\pi$ branching ratios. For the same reason, the
penguin annihilation amplitudes, which involve the $\pi$-$K$ or
$\pi$-$\pi$ time-like form factor, decrease. The magnitude of the
resultant direct CP asymmetries of the $B\to\pi K$, $\pi\pi$
decays, which is not compensated by the overall decay constant
$f_B$, then becomes smaller than in \cite{KS} as shown in the
column LO of Table~\ref{cp2}. The smaller $B^0\to\pi^\mp K^\pm$
direct CP asymmetry is in better agreement with the data, implying
that the data could be covered by the theoretical uncertainty at
LO of PQCD.

All the above nonperturbative inputs suffer uncertainties, and it
is necessary to investigate how these uncertainties propagate into
the predictions for nonleptonic $B$ meson decays. Here we shall
constrain the shape parameter $\omega_B$ and the Gegenbauer
coefficients of the twist-2 pion distribution amplitude
$\phi_\pi^A$ using the experimental error of the semileptonic
decay $B\to\pi l\nu$. The sufficient uncertainties will be
assigned to the Gegenbauer coefficients of the twist-2 kaon
distribution amplitude $\phi_K^A$. The other inputs, such as the
$B$ meson decay constant, the twist-3 distribution amplitudes, and
the chiral scale associated with the pion and the kaon will be
fixed. On one hand, the considered sources of theoretical
uncertainties have been representative enough. On the other hand,
it is impossible to constrain all the inputs with the currently
available data.

The spectrum of the semileptonic decay $B\to\pi l\nu$ in the
lepton invariant mass $q^2$ has been measured \cite{CLEO}:
\begin{eqnarray}
\frac{\int_0^8 (d\Gamma/d q^2) dq^2}{\Gamma_{\rm total}}\, =\,
(0.43\pm 0.11)\times 10^{-4}\;,
\end{eqnarray}
with the total decay rate $\Gamma_{\rm total}=(4.29\pm 0.04)\times
10^{-13}$ GeV \cite{PDG}. Assuming that the above error is uniform
in the region $0 < q^2 < 8$ GeV$^2$, we derive the uncertainty
$\Delta$ of $(d\Gamma/d q^2) |_{q^2 =0}$ by solving the equation
$8\Delta=0.11\times 10^{-4}\,\Gamma_{\rm total}$, where we take only
the central value of $\Gamma_{\rm total}$ for simplicity. With the
allowed range of $|V_{ub}|=(3.67 \pm 0.47)\times 10^{-3}$
\cite{PDG}, $\Delta$ is translated into the uncertainty of the
$B\to \pi$ form factor,
\begin{eqnarray}
F_+^{B \pi}(0)\, =\, 0.24\pm 0.05\;,\label{bpi}
\end{eqnarray}
whose central value comes from our choice of the inputs. Equation
(\ref{bpi}) is consistent with $0.23\pm 0.04$ extracted in
\cite{LR03} from a global fit to the above CLEO data, lattice QCD
results of $F_+^{B\pi}(q^2)$, etc. A numerical analysis indicates
that $F_+^{B \pi}(0)$ is more sensitive to $\omega_B$ than to the
Gegenbauer coefficients of $\phi_\pi^A$.

Therefore, we propose
\begin{enumerate}
\item the shape parameters for the distribution amplitudes,
\begin{eqnarray}
& &\omega_B\, =\,  (0.40\pm 0.04)\; {\rm GeV}\;,\;\;\;\;
a_2^\pi\, =\, 0.115\pm 0.115\;,\;\;\;\;a_4^\pi\, =\, -0.015\;,
\nonumber\\
& &a_1^K\, =\, 0.17\pm 0.17\;\;,\;\;\;\;
a_2^K\, =\, 0.115\pm 0.115\;,\;\;\;\; a_4^K\, =\, -0.015\;,
\label{sha}
\end{eqnarray}
that is, the Gegenbauer coefficients can vary by 100\%.
%The Gegenbauer coefficients of the twist-2 kaon distribution amplitude
%$\phi_K^A$ have been assumed to vary in a similar way.
We do not consider the uncertainty from the coefficients $a_4^\pi$
and $a_4^K$, to which our predictions are insensitive. Note that
the first Gegenbauer coefficients $a_1^K\approx 0.10\pm 0.12$ and
$a_1^K\approx 0.05\pm 0.02$ have been found to be smaller in
\cite{BL04} and \cite{KMM04}, respectively. A hint on the effect
from the evolution of the meson distribution amplitudes from $1/b$
down to the cutoff $\mu_0$ (see Sec. III) can also be obtained
through the above variation of the Gegenbauer coefficients.

\item the CKM matrix elements,
\begin{eqnarray}
& &V_{ud}\, =\, 0.9734\;,\;\;\;\; V_{us}\, =\, 0.2200\;,\;\;\;\;
|V_{ub}|\, =\, (3.67 \pm 0.47)\times 10^{-3}\;,\nonumber\\
& &V_{cd}\, =\, -0.224\;,\;\;\;\; V_{cs}\, =\, 0.996\;,\;\;\;\;
V_{cb}\, =\, 0.0413\;,
\end{eqnarray}
where we consider only the representative source of theoretical
uncertainties from $|V_{ub}|$ \cite{PDG}. This source is essential
for estimating the uncertainty of the predicted direct CP
asymmetries. $V_{cb}=(41.3\pm 1.5)\times 10^{-3}$ \cite{PDG} has a
smaller uncertainty, and the other matrix elements have been known
more precisely. The unitarity condition $V_{tb}V^*_{ts(d)} =
-V_{ub}V^*_{us(d)} -V_{cb}V^*_{cs(d)}$ is then employed as
evaluating the penguin contributions.

\item the weak phases,
\begin{eqnarray}
\phi_1\, =\, 21.6^\circ \;,\;\;\;\; \phi_3\, =\, (70\pm 30)^\circ\;,
\end{eqnarray}
where the range of the well-measured $\phi_1$ with
$\sin(2\phi_1)=0.685\pm 0.032$ \cite{Abe} has been neglected, and
the range of $\phi_3$ is hinted by the determinations
\cite{Abe,Smith},
\begin{eqnarray}
\phi_3&=&68^{+14}_{-15} \pm 13 \pm 11\;\;  {\rm (Belle,
Dalitz)}\;,\nonumber\\
& & 70 \pm 31^{+12+14}_{-10-11}\;\; {\rm (BaBar,
Dalitz)}\;,\nonumber\\
& &63^{+15}_{-13}\;\;  {\rm (CKMfitter)}\;,\nonumber\\
& & 64\pm 18  \;\;      {\rm (UTfit)}\;.
\end{eqnarray}

\end{enumerate}

We fix the other parameters, such as the meson decay constants
$f_{B}= 210$ MeV, $f_{K}= 160$ MeV, $f_{\pi}= 130$ MeV, the meson
masses $m_B = 5.28$ GeV, $m_K=0.49$ GeV, $m_\pi=0.14$ GeV, the
charm quark mass $m_c=1.5$ GeV, and the $B$ meson lifetimes
%$\tau_{B^0}=1.542\times 10^{-12}$ sec, $\tau_{B^\pm}= 1.674\times
%10^{-12}$ sec \cite{PDG}.
$\tau_{B^0}=1.528\times 10^{-12}$ sec, $\tau_{B^\pm}= 1.643\times
10^{-12}$ sec \cite{HFAG}.  We also fix the chiral scales $m_{0\pi}=
1.3$ GeV and $m_{0K} = 1.7$ GeV, where the value of $m_{0\pi}$
($m_{0K}$) is close to that (larger than $1.25\pm 0.15$ GeV)
obtained in the recent sum-rule analysis \cite{HZW05}. The
resultant $B\to \pi, K$ transition form factors,
\begin{eqnarray}
F_+^{B\pi}(0)\, =\, 0.24^{+0.05}_{-0.04}\;,\;\;\;\;
F_+^{BK}(0)\, =\, 0.36^{+0.09}_{-0.07} \;,
\end{eqnarray}
respect Eq.~(\ref{bpi}) from the measurement, and are consistent
with the estimation from light-cone sum rules \cite{KMM04}. If
further including the variation of $m_{0K}$ as a source of
theoretical uncertainties, we just enlarge the range of the
$B\to\pi K$ branching ratios, but not of the other quantities. We
have tested the dependence of our predictions on the cutoff
$\mu_0$, which is found to be weak.

The above inputs lead to Tables~\ref{br2}-\ref{mixcp}, where the
theoretical uncertainties are displayed only in the columns +NLO.
The errors (not) in the parentheses represent those from (all
sources) the first source of uncertainties. It indicates that the
nonperturbative inputs, i.e., the first source, contribute to the
theoretical uncertainties more dominantly in the $B\to\pi K$
decays than in the $B\to\pi \pi$ decays, because the former depend
on the less controllable parameters associated with the kaon. We
also observe that $A_{CP}(B^0\to \pi^\mp K^\pm)$ and
$A_{CP}(B^\pm\to \pi^0 K^\pm)$ always increase or decrease
simultaneously, when varying the nonperturbative inputs. Hence,
the $B\to\pi K$ puzzle can not be resolved by tuning these
parameters. After including the uncertainties, the predicted
$B^0\to\pi^0\pi^0$ branching ratio and mixing-induced CP asymmetry
$S_{\pi^0K_S}$ are still far from the data.

%%%%%%%%%%%%%%%%%%%%%   Figure 4  %%%%%%%%%%%%%%%%%%%%%%%%%%%%%%%
\begin{figure}[tb]
\begin{center}
\includegraphics[scale=0.8]{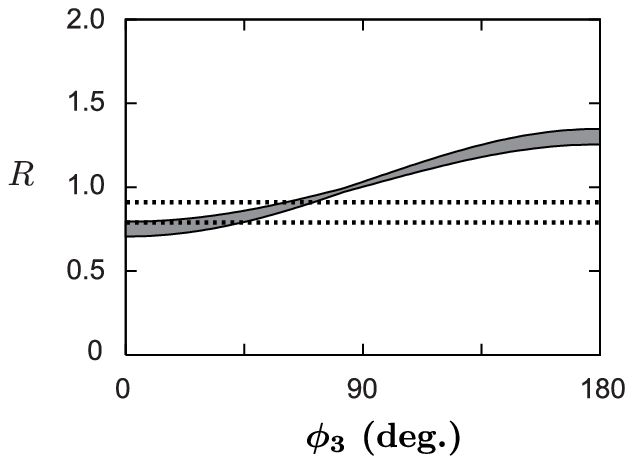}
 \includegraphics[scale=0.8]{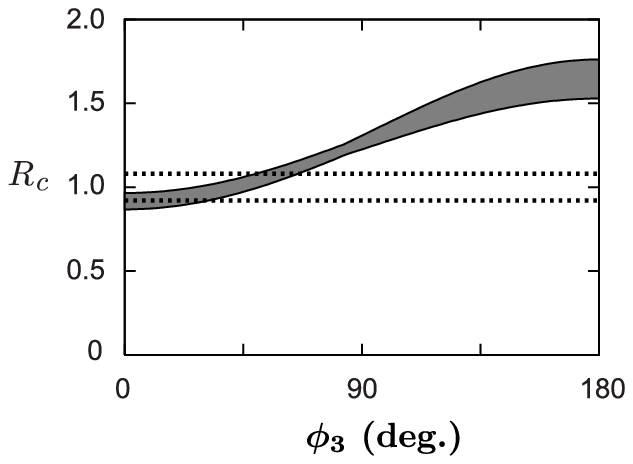}
 \includegraphics[scale=0.8]{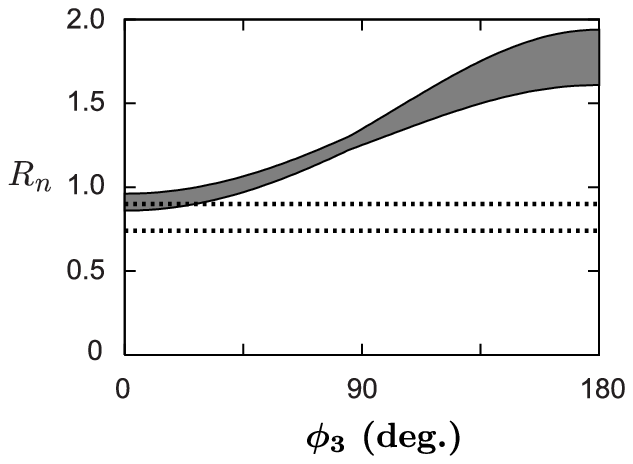}
\caption{Dependence of $R$, $R_c$, and $R_n$ on $\phi_3$ from NLO
PQCD with the bands representing the theoretical uncertainty. The
two dashed lines represent $1\sigma$ bounds from the data.
}\label{three}
\end{center}
\end{figure}
%%%%%%%%%%%%%%%%%%%%%%%%%%%%%%%%%%%%%%%%%%%%%%%%%%%%%%%%%%%%%%%%%

A more transparent comparison between the predictions and the data
is made by considering the ratios of the branching ratios. The
following three ratios of the $B\to\pi K$ branching ratios have
been widely studied in the literature,
\begin{eqnarray}
R&=&
\frac{B(B^0\to\pi^\mp K^\pm)}{B(B^\pm\to\pi^\pm K^0)}
\frac{\tau_{B^+}}{\tau_{B^0}}\, =\,
0.85\pm 0.06
%0.82\pm 0.06
\;,\nonumber\\
R_c&=&
2\, \frac{B(B^\pm\to\pi^0 K^\pm)}{B(B^\pm\to\pi^\pm K^0)}\, =\,
1.00\pm 0.08
\;,\nonumber\\
R_n&=&
\frac{1}{2}\frac{B(B^0\to\pi^\mp K^\pm)}{B(B^0\to\pi^0 K^0)}\, =\,
0.82\pm 0.08
%0.79\pm 0.08
\;,\label{3R}
\end{eqnarray}
whose values are quoted from \cite{HFAG}. We have confirmed that
these ratios depend on the nonperturbative inputs weakly.
Therefore, their deviation from the standard-model predictions
could reveal a signal of new physics, such as a large electroweak
penguin amplitude. Table~\ref{br2} shows that for
$\phi_3=70^\circ$, the ratio $R$ increases slightly from 0.90 to
0.92, when the NLO Wilson coefficients are adopted, beyond which
the various NLO corrections do not change $R$ much. The ratio
$R_c$ ($R_n$) decreases from 1.20 (1.25) to 1.14 (1.14), when the
NLO Wilson coefficients are adopted, and settles down at this
value as indicated by the column +NLO. The different types of NLO
corrections cause only small fluctuations. Comparing the columns
LO and +NLO, the consistency between the PQCD predictions and the
data has been improved.

Varying the weak phase $\phi_3$ and the inputs, we find that the
PQCD predictions for $R$ and $R_n$ are in good agreement with the
data in Eq.~(\ref{3R}), which is obvious from Fig.~\ref{three}.
However, the predictions for $R_n$ exhibit a tendency of
overshooting the data, which is attributed to the smaller PQCD
results for the $B^0\to\pi^0 K^0$ branching ratio. A smaller
Gegenbauer coefficient $a_2^\pi$ of $\phi_\pi^A$ enhances $R_n$.
That is, when using the updated pion distribution amplitudes from
\cite{Ball:2004ye}, the consistency of the predictions for $R_n$
with the data deteriorates. A smaller $\omega_B$ enhances $R_n$.
This is the reason we do not lower $\omega_B$ in order to
compensate the reduction from the smaller $a_2^\pi$. Note that
$m_{0K}$ has an effect on the electroweak penguin amplitude, i.e.,
on the $B^0\to\pi^0 K^0$ branching ratio. Hence, we have also
studied the dependence of $R_n$ on the chiral scale $m_{0K}$. A
smaller $m_0^K$ indeed reduces $R_n$, but does not help much:
choosing $m_{0K}=1.3$ GeV causes only few percent reduction of
$R_n$. It has been known that the $B^0\to\pi^0 K^0$ branching
ratio can be significantly increased by rotating the electroweak
penguin amplitude $P'_{ew}$ away from the penguin amplitude $P'$
(their values in Table~\ref{tp} are roughly parallel to each
other). Therefore, we can not rule out the possibility that
$P'_{ew}$ acquires an additional phase from new physics effects
\cite{BFRPR,HNS,BCLL}. However, our theoretical uncertainty is
representative, and the actual uncertainty could be larger, such
that the discrepancy is not serious at this moment. We do not
discuss the ratios relevant to the $B\to\pi\pi$ decays, because
the PQCD predictions for the $B^0\to\pi^0\pi^0$ branching ratio
are currently far below the measured values.

\section{CONCLUSION}

The LO PQCD has correctly predicted the direct CP asymmetry
$A_{CP}(B^0\to \pi^\mp K^\pm)$, but failed to explain another one
$A_{CP}(B^\pm\to \pi^0K^\pm)$ \cite{KLS}. Phenomenologically, the
substantial difference between $A_{CP}(B^\pm\to \pi^0K^\pm)$ and
$A_{CP}(B^0\to \pi^\mp K^\pm)$ has led to the conjecture of new
physics \cite{BFRPR,Y03}. However, the difference can also be
attributed to a large color-suppressed tree amplitude $C'$ as
pointed out in \cite{Charng2}. Theoretically, an examination of
NLO effects is always demanded for a systematic approach like
PQCD. Since $C'$ itself is a subdominant contribution, it is
easily affected by subleading corrections. Hence, before claiming
a new physics signal in the $B\to \pi K$ data, one should at least
check whether the NLO effects could enhance $C'$ sufficiently.
This is one of our motivations to perform the NLO calculation in
PQCD for the $B\to\pi K$, $\pi\pi$ decays here. Another motivation
comes from the mixing-induced CP asymmetries in the
penguin-dominated modes, some of which also depend on the
color-suppressed tree amplitudes. To estimate the deviation of
$S_{\pi^0K_S}$ from $S_{c\bar cs}$ within the standard model, one
must compute $C'$ reliably.

In this paper we have calculated the NLO corrections to the
$B\to\pi K$, $\pi\pi$ decays from the vertex corrections, the
quark loops, and the magnetic penguin in the PQCD approach. The
results for the branching ratios and CP asymmetries in the NDR
scheme have been presented in Tables~\ref{br2}-\ref{mixcp}, and
discussed in Sec.~III. It has been shown that the corrections from
the quark loops and from the magnetic penguin come with opposite
signs, and sum to about 10\% of the LO penguin amplitudes. Their
effect is to reduce the $B\to\pi K$ branching ratios, to which the
penguin contribution is dominant, by about 20\%. They have a minor
influence on the $B\to\pi\pi$ branching ratios, and CP
asymmetries. The vertex corrections play an important role in
modifying direct CP asymmetries, especially those of the
$B^\pm\to\pi^0 K^\pm$, $B^0\to\pi^0K^0$, and $B^0\to\pi^0\pi^0$
modes, by increasing the color-suppressed tree amplitudes few
times. The larger color-suppressed tree amplitude leads to nearly
vanishing $A_{CP}(B^\pm\to \pi^0K^\pm)$, resolving the $B\to\pi K$
puzzle within the standard model. Our analysis has also confirmed
that the NLO corrections are under control in PQCD.

The NLO corrections, though increasing the color-suppressed tree
amplitudes significantly, are not enough to enhance the
$B^0\to\pi^0\pi^0$ branching ratio to the measured value. A much
larger amplitude ratio $|C/T|\sim 0.8$ must be obtained in order
to resolve this puzzle \cite{Charng2}. Nevertheless, the NLO
corrections do improve the consistency of our predictions with the
data: the predicted $B^0\to\pi^\pm\pi^\mp$ ($B^0\to\pi^0\pi^0$)
branching ratio decreases (increases). Viewing the consistency of
the PQCD predictions with the tiny measured $B^0\to K^0\overline
K^0$ and $B^0\to\rho^0\rho^0$ branching ratios, we think that our
NLO results for the $B\to\pi\pi$ decays are reasonable. In SCET
\cite{BPRS}, the large $|C/T|$ comes from a fit to the data,
instead of from an explicit evaluation of the amplitudes. The
amplitude $C$ was indeed found to be increased in SCET by the NLO
jet function (the short-distance coefficient from matching
SCET$_{\rm I}$ to SCET$_{\rm II}$) \cite{BY05}, if the parameter
set ''S4" in QCDF \cite{BBNS} was adopted. The large measured
$B^0\to\pi^0\pi^0$ branching ratio was then explained. However, we
emphasize again that the same analysis should be applied to the
$B\to\rho\rho$ decays for a check. Hence, the $B\to\pi\pi$ puzzle
still requires more investigation.

The tendency of overshooting the observed ratio $R_n$ has implied
a possible new-physics phase associated with the electroweak
penguin amplitude $P'_{ew}$. This additional phase can render
$P'_{ew}$ orthogonal to the penguin amplitude, and enhance the
$B^0\to\pi^0 K^0$ branching ratio. We have also computed the
deviation $\Delta S_{\pi^0 K_S}$ of the mixing-induced CP
asymmetry, and found that the NLO effects push it toward the even
larger positive value. Therefore, it is difficult to understand
the observed negative deviation without physics beyond the
standard model.

\vskip 1.0cm

We thank  T. Browder, H.Y. Cheng, C.K. Chua, W.S. Hou, Y.Y. Keum,
M. Nagashima, A. Soddu, and A. Soni for useful discussions. This
work was supported by the National Science Council of R.O.C. under
Grant No. NSC-94-2112-M-001-001, and by the Grants-in-aid from the
Ministry of Education, Culture, Sports, Science and Technology,
Japan under Grant No. 14046201. HNL acknowledges the hospitality
of Department of Physics, Hawaii University, where this work was
initiated.

\appendix*

\section{FACTORIZATION FORMULAS}

We first define the kinematics for the $B\to M_2M_3$ decay, where
$M_2$ ($M_3$) denotes the light pseudo-scalar meson involved in
the $B$ meson transition (emitted from the weak vertex). In the
rest frame of the $B$ meson, the $B$ ($M_2$, $M_3$) meson momentum
$P_1$ ($P_2$, $P_3$), and the corresponding spectator quark
momentum $k_1$ ($k_2$, $k_3$) are taken, in the light-cone
coordinates, as
\begin{eqnarray}
  P_1\, =\, \frac{m_B}{{\sqrt 2}}(1,1,{\bf 0}_T)
  \;, \;\;\;\;\;\;\;\;\;\;\;\;\;\;\;\;\;\;\;
  k_1\, =\, (0, x_1P_1^-,{\bf k}_{1T})\;,\nonumber\\
  P_2\, =\, \frac{m_B}{{\sqrt 2}}(1,0,{\bf 0}_T)
  \;, \;\;\;\;\;\;\;\;\;\;\;\;\;\;\;\;\;\;\;
  k_2\, =\, (x_2P_2^+,0, {\bf k}_{2T})\;,\nonumber\\
  P_3\, =\, \frac{m_B}{{\sqrt 2}}(0,1,{\bf 0}_T)
  \;, \;\;\;\;\;\;\;\;\;\;\;\;\;\;\;\;\;\;\;
  k_3\, =\, (0, x_3P_3^-, {\bf k}_{3T})\;,
\end{eqnarray}
where the light meson masses have been neglected. We also define
the ratio $r_2=m_{02}/m_B$ ($r_3=m_{03}/m_B$)
associated with the meson $M_2$ ($M_3$), $m_{02}$ ($m_{03}$) being
the chiral scale.

The factorization formulas for the $B \to M_2 M_3$ decay
amplitudes appearing in Tables~\ref{amp} and \ref{ampp} are
collected below:
\begin{eqnarray}
F_{e4}(a) &=& 16  \pi C_F m_B^2 \int_0^1 dx_1 dx_2 \int_0^{\infty}
b_1db_1\, b_2db_2\, \phi_B(x_1,b_1)
\nonumber \\
& &\times \bigg\{ \left[
(1+x_2)\phi_{2}^A(\overline{x_2})+r_{2}(1-2x_2) \left(
\phi_{2}^P(\overline{x_2})-\phi_{2}^T(\overline{x_2}) \right)
\right]
E_{e}(t) h_{e}(A,B,b_1,b_2,x_2)
\nonumber\\
& &\;\;\;\;\;\; +2r_{2} \phi_{2}^P(\overline{x_2})E_{e}(t')
h_{e}(A',B',b_2,b_1,x_1) \bigg\} \;,
\\
F_{e6}(a) &=&
 32 \pi C_F m_B^2 \int_0^1 dx_1 dx_2 \int_0^{\infty}
b_1db_1\, b_2db_2\, \phi_B(x_1,b_1)
\nonumber \\
& &\times \bigg\{  r_3 \left[ \phi_2^A(\overline{x_2})+r_2 x_2
\left( \phi_2^P(\overline{x_2}) + \phi_2^T(\overline{x_2}) \right)
+2 r_2 \phi_2^P(\overline{x_2})
 \right]
E_{e}(t) h_{e}(A,B,b_1,b_2,x_2)
\nonumber\\
& &\;\;\;\;\;\; + 2 r_2 r_3
 \phi_2^P(\overline{x_2})
E_{e}(t') h_{e}(A',B',b_2,b_1,x_1) \bigg\}\;,
\\
F_{a4}(a) &=& 16 \pi C_F m_B^2 \int_0^1 dx_2 dx_3 \int_0^{\infty}
b_2db_2\, b_3db_3\,
\nonumber \\
& &\times \bigg\{ \left[
x_3\phi_{2}^A(\overline{x_2})\phi_{3}^A(\overline{x_3}) + 2r_{2}
r_{3} \phi_{2}^P(\overline{x_2}) \left\{ \left(
\phi_{3}^P(\overline{x_3}) + \phi_{3}^T(\overline{x_3}) \right) +
x_3\left( \phi_{3}^P(\overline{x_3}) - \phi_{3}^T(\overline{x_3})
\right) \right\} \right]
\nonumber \\
& &\;\;\;\;\;\;\times E_{a}(t) h_{e}(A,B,b_2,b_3,x_3)
\nonumber\\
& &\;\;\;\;\;\;
 - \left[
(1-x_2)\phi_{2}^A(\overline{x_2})\phi_{3}^A(\overline{x_3}) -
2r_{2} r_{3} \left\{ -2\phi_{2}^P(\overline{x_2}) + x_2\left(
\phi_{2}^P(\overline{x_2})+\phi_{2}^T(\overline{x_2}) \right)
\right\} \phi_{3}^P(\overline{x_3}) \right]
\nonumber \\
& &\;\;\;\;\;\;\times E_{a}(t') h_{e}(A',B',b_3,b_2,x_2)
\bigg\}\;,
\\
F_{a6}(a) &=& 32 \pi C_F m_B^2 \int_0^1 dx_2 dx_3 \int_0^{\infty}
b_2db_2\, b_3db_3\,
\nonumber \\
& &\times \bigg\{ \left[ 2 r_{2}
\phi_{2}^P(\overline{x_2})\phi_{3}^A(\overline{x_3}) + r_{3}x_3
\phi_{2}^A(\overline{x_2}) \left( \phi_{3}^P(\overline{x_3}) +
\phi_{3}^T(\overline{x_3}) \right) \right] E_{a}(t)
h_{e}(A,B,b_2,b_3,x_3)
\nonumber\\
& &\;\;\;\;\;\;
 + \left[
r_{2} (1-x_2)\left(
\phi_{2}^P(\overline{x_2})-\phi_{2}^T(\overline{x_2})
         \right)\phi_{3}^A(\overline{x_3})
+
 2 r_{3} \phi_{2}^A(\overline{x_2}) \phi_{3}^P(\overline{x_3}) \right]
\nonumber\\
& & \;\;\;\;\;\;\times E_{a}(t') h_{e}(A',B',b_3,b_2,x_2)
\bigg\}\;,
\\
{\cal M}_{e4}(a') &=& 32 \pi C_F
\frac{\sqrt{2N_c}}{N_c}m_B^2\int_0^1 dx_1dx_2dx_3 \int_0^{\infty}
b_1 db_1\, b_3 db_3\,\phi_B(x_1,b_1) \phi_{3}^A(\overline{x_3})
\nonumber \\
& &\times \bigg\{ \left[(1-x_3)\phi_{2}^A(\overline{x_2}) -r_{2}
x_2 \left( \phi^P_{K}(\overline{x_2}) +
        \phi^T_{K}(\overline{x_2}) \right) \right]
E'_{e}(t) h_n(A,B,b_1,b_3)
\nonumber \\
& &\;\;\;\;\;\; + \left[-(x_2+x_3)\phi_{2}^A(\overline{x_2})
+r_{2} x_2 \left(
\phi_{2}^P(\overline{x_2})-\phi_{2}^T(\overline{x_2})
       \right) \right]
E'_{e}(t') h_n(A',B',b_1,b_3) \bigg\}\;,\label{me4}
\\
%%
%{\cal M}_{e5}(a') &=& 32 \pi C_F
%\frac{\sqrt{2N_c}}{N_c}m_B^2\int_0^1 dx_1dx_2dx_3 \int_0^{\infty}
%b_1 db_1\, b_3 db_3\,\phi_B(x_1,b_1) \phi_{3}^A(\overline{x_3})
%\nonumber \\
%& &\times \bigg\{ \left[- (1+x_2-x_3)\phi_{2}^A(\overline{x_2})
%+r_{2} x_2 \left( \phi^P_{2}(\overline{x_2}) -
%        \phi^T_{2}(\overline{x_2}) \right) \right]
%E'_{e}(t) h_n(A,B,b_1,b_3)
%\nonumber \\
%& &\;\;\;\;\;\; + \left[x_3 \phi_{2}^A(\overline{x_2}) -r_{2} x_2
%\left(  \phi_{2}^P(\overline{x_2})+\phi_{2}^T(\overline{x_2})
%       \right) \right]
%E'_{e}(t') h_n(A',B',b_1,b_3) \bigg\}\;,
%\\
%
{\cal M}_{e6}(a') &=& 32  \pi C_F \frac{\sqrt{2N_c}}{N_c}m_B^2\,
r_{3} \int_0^1 dx_1dx_2dx_3\int_0^{\infty} b_1 db_1\, b_3
db_3\,\phi_B(x_1,b_1)
\nonumber \\
& &\times \bigg\{ \left[ (1-x_3)\phi_{2}^A(\overline{x_2}) \left(
\phi_{3}^P(\overline{x_3}) - \phi_{3}^T(\overline{x_3}) \right)
\right.\nonumber \\
& &\;\;\;\;\;\;+r_{2}(1-x_3)\left(
\phi_{2}^P(\overline{x_2})+\phi_{2}^T(\overline{x_2}) \right)
\left( \phi_{3}^P(\overline{x_3}) - \phi_{3}^T(\overline{x_3})
\right)\nonumber\\
& &\;\;\;\;\;\;\left. + r_{2} x_2 \left(
\phi_{2}^P(\overline{x_2})-\phi_{2}^T(\overline{x_2}) \right)
\left( \phi_{3}^P(\overline{x_3}) + \phi_{3}^T(\overline{x_3})
\right) \right] E'_{e}(t) h_n(A,B,b_1,b_3)
\nonumber \\
& & - \left[ x_3\phi_{2}^A(\overline{x_2}) \left(
\phi_{3}^P(\overline{x_3}) + \phi_{3}^T(\overline{x_3}) \right)
+ r_{2}x_3 \left(
\phi_{2}^P(\overline{x_2})+\phi_{2}^T(\overline{x_2}) \right)
\left( \phi_{3}^P(\overline{x_3}) + \phi_{3}^T(\overline{x_3})
\right)
\right.\nonumber \\
& &\;\;\;\;\;\;\left. +r_{2} x_2 \left(
\phi_{2}^P(\overline{x_2})-\phi_{2}^T(\overline{x_2})
       \right)
\left( \phi_{3}^P(\overline{x_3}) - \phi_{3}^T(\overline{x_3})
\right) \right] E'_{e}(t') h_n(A',B',b_1,b_3) \bigg\}\;,
\\
{\cal M}_{a4}(a') &=& 32 \pi C_F
\frac{\sqrt{2N_c}}{N_c}m_B^2\int_0^1 dx_1dx_2dx_3 \int_0^{\infty}
b_1 db_1\, b_3 db_3\,\phi_B(x_1,b_1)
\nonumber \\
& &\times \bigg\{ \left[
(1-x_2)\phi_{2}^A(\overline{x_2})\phi_{3}^A(\overline{x_3})
+ r_{2} r_{3} \left\{ (1-x_2)\left(
\phi_{2}^P(\overline{x_2})+\phi_{2}^T(\overline{x_2}) \right)
\left( \phi_{3}^P(\overline{x_3}) - \phi_{3}^T(\overline{x_3})
\right)
\right.\right.\nonumber \\
& &\;\;\;\;\;\;\left.\left. +x_3\left(
\phi_{2}^P(\overline{x_2})-\phi_{2}^T(\overline{x_2}) \right)
\left( \phi_{3}^P(\overline{x_3}) + \phi_{3}^T(\overline{x_3})
\right) \right\} \right]
E'_{a}(t) h_n(A,B,b_3,b_1)
\nonumber \\
& & - \left[
x_3\phi_{2}^A(\overline{x_2})\phi_{3}^A(\overline{x_3})
+ r_{2} r_{3} \left\{
4\phi_{2}^P(\overline{x_2})\phi_{3}^P(\overline{x_3})
\right.\right.\nonumber \\
& &\;\;\;\;\;\;-(1-x_3) \left(
\phi_{2}^P(\overline{x_2})+\phi_{2}^T(\overline{x_2}) \right)
\left( \phi_{3}^P(\overline{x_3}) - \phi_{3}^T(\overline{x_3})
\right)\nonumber\\
& &\;\;\;\;\;\;\left.\left.
 - x_2\left(  \phi_{2}^P(\overline{x_2})-\phi_{2}^T(\overline{x_2}) \right)
\left( \phi_{3}^P(\overline{x_3}) + \phi_{3}^T(\overline{x_3})
\right) \right\} \right]
E'_{a}(t') h_n(A',B',b_3,b_1) \bigg\}\;,
\\
%%
%{\cal M}_{a5}(a') &=& 32 \pi C_F
%\frac{\sqrt{2N_c}}{N_c}m_B^2\int_0^1 dx_1dx_2dx_3 \int_0^{\infty}
%b_1 db_1\, b_3 db_3\,\phi_B(x_1,b_1)
%\nonumber \\
%& &\times \bigg\{ \left[
%x_3\phi_{2}^A(\overline{x_2})\phi_{3}^A(\overline{x_3})
%+ r_{2} r_{3} \left\{ (1-x_2)\left(
%\phi_{2}^P(\overline{x_2})-\phi_{2}^T(\overline{x_2}) \right)
%\left( \phi_{3}^P(\overline{x_3}) + \phi_{3}^T(\overline{x_3})
%\right)
%\right.\right.\nonumber \\
%& &\;\;\;\;\;\;\left.\left. +x_3\left(
%\phi_{2}^P(\overline{x_2})+\phi_{2}^T(\overline{x_2})
%      \right)
%\left( \phi_{3}^P(\overline{x_3}) - \phi_{3}^T(\overline{x_3})
%\right) \right\} \right]
%E'_{a}(t) h_n(A,B,b_3,b_1)
%\nonumber \\
%& & - \left[
%(1-x_2)\phi_{2}^A(\overline{x_2})\phi_{3}^A(\overline{x_3})
%\right.\nonumber\\
%& &\left. \;\;\;\;\;\; + r_{2} r_{3} \left\{
%4\phi_{2}^P(\overline{x_2})\phi_{3}^P(\overline{x_3}) -(1-x_3)
%\left(  \phi_{2}^P(\overline{x_2})-\phi_{2}^T(\overline{x_2})
%\right) \left( \phi_{3}^P(\overline{x_3}) +
%\phi_{3}^T(\overline{x_3}) \right)
%\right.\right.\nonumber \\
%& &\;\;\;\;\;\;\left.\left.
% - x_2\left(  \phi_{2}^P(\overline{x_2})+\phi_{2}^T(\overline{x_2}) \right)
%\left( \phi_{3}^P(\overline{x_3}) - \phi_{3}^T(\overline{x_3})
%\right) \right\} \right]
%E'_{a}(t') h_n(A',B',b_3,b_1) \bigg\}\;,
%\\
%
{\cal M}_{a6}(a') &=& 32 \pi C_F
\frac{\sqrt{2N_c}}{N_c}m_B^2\int_0^1 dx_1dx_2dx_3 \int_0^{\infty}
b_1 db_1\, b_3 db_3\,\phi_B(x_1,b_1)
\nonumber \\
& &\times \bigg\{ \Big[ -r_{2} (1-x_2) \left(
\phi^P_{2}(\overline{x_2}) + \phi^T_{2}(\overline{x_2}) \right)
\phi_{3}^A(\overline{x_3})
\nonumber\\
& & \;\;\;\;\;\; + r_{3} x_3 \phi^A_{2}(\overline{x_2}) \left(
\phi_{3}^P(\overline{x_3}) - \phi_{3}^T(\overline{x_3}) \right)
\Big] E'_{a}(t) h_n(A,B,b_3,b_1)
\nonumber \\
& & - \Big[ r_{2} (1+x_2) \left( \phi^P_{2}(\overline{x_2}) +
\phi^T_{2}(\overline{x_2}) \right) \phi_{3}^A(\overline{x_3})
\nonumber\\
& & \;\;\;\;\;\; + r_{3} (-2+x_3) \phi^A_{2}(\overline{x_2})
\left( \phi_{3}^P(\overline{x_3}) -
\phi_{3}^T(\overline{x_3})\right)
 \Big]
E'_{a}(t') h_n(A',B',b_3,b_1) \bigg\}\;,
\end{eqnarray}
where we have adopted the notations $\overline{x_2} \equiv 1-x_2$
and $\overline{x_3} \equiv 1-x_3$, and ignored the mass difference
between $m_B$ and $m_b$.  $F_{eKi}$ and ${\cal M}_{eKi}$ are
obtained by choosing $M_2$ ($M_3$) to be the kaon (pion) in
$F_{ei}$ and ${\cal M}_{ei}$, respectively.

The invariant masses $A$, $B$, $A'$, and $B'$ of the virtual quarks
and gluons involved in the above hard kernels are functions of
$x_1$, $x_2$ and $x_3$, as in Table~\ref{tab:abcphik}.
%
%%%%%%%%%%%%%%%%%%%%%   Table VIII  %%%%%%%%%%%%%%%%%%%%%%%%%%%%%
\begin{table}[bt]
\begin{center}
\begin{tabular}{c|cccc}
\hline\hline
 & $A$ & $B$ & $A'$ & $B'$ \\
\hline
$F_{e4,e6}$ & $\sqrt{x_2}m_B$ & $\sqrt{x_1x_2}m_B$
    & $\sqrt{x_1}m_B$ & $\sqrt{x_1x_2}m_B$
    \\
${\cal M}_{e4,e5,e6}$ & $i \sqrt{x_2(\overline{x_3}-x_1)}m_B$ &
      $\sqrt{x_1x_2}m_B$
      & $\sqrt{x_2(x_1-x_3)}m_B$
      & $\sqrt{x_1x_2}m_B$
    \\
$F_{a4,a6}$ & $i \sqrt{x_3}m_B$ & $ i \sqrt{\overline{x_2}x_3}m_B$
      & $ i \sqrt{\overline{x_2}}m_B$ & $ i \sqrt{\overline{x_2}x_3}m_B$
    \\
${\cal M}_{a4,a5,a6}$ & $\sqrt{(x_1-x_3)\overline{x_2}}m_B$ &
      $ i \sqrt{\overline{x_2}x_3}m_B$
      & $\sqrt{1-x_2(\overline{x_3}-x_1)}m_B$ & $ i
 \sqrt{\overline{x_2}x_3}m_B$
    \\
\hline\hline
\end{tabular}
\caption{The invariant masses $A$, $B$, $A'$, and $B'$ in the hard
kernels.} \label{tab:abcphik}
\end{center}
\end{table}
%%%%%%%%%%%%%%%%%%%%%%%%%%%%%%%%%%%%%%%%%%%%%%%%%%%%%%%%%%%%%%%%%
%
The hard scales are chosen as
\begin{eqnarray}
t &=&{\rm max}(\sqrt{|A^2|},\, \sqrt{|B^2|},\, 1/b_i)
\;,\nonumber\\
t' &=&{\rm max}(\sqrt{|A'^2|},\, \sqrt{|B'^2|},\, 1/b_i) \;,
\label{ttp}
\end{eqnarray}
with the index $i=1,2$ for $F_{e4,e6}$, $i=2,3$ for $F_{a4,a6}$,
and $i=1,3$ for the nonfactorizable amplitudes.

The evolution factors $E_{e}^{(\prime)}$ and $E_{a}^{(\prime)}$
are given by
\begin{eqnarray}
E_{e}(t)&=&\alpha_s(t)\,a(t)\exp[-S_B(t)-S_{2}(t)]
\;,\nonumber\\
E_{a}(t)&=&\alpha_s(t)\,a(t)\exp[-S_{2}(t)-S_{3}(t)]
\;,\nonumber\\
E'_{e}(t)&=&\alpha_s(t)\,a'(t)\exp[-S_B(t)-S_{2}(t)-S_{3}(t)]|_{b_2=b_1}
\;,\nonumber\\
E'_{a}(t)&=&\alpha_s(t)\,a'(t)\exp[-S_B(t)-S_{2}(t)-S_{3}(t)]|_{b_2=b_3}
\;,
\end{eqnarray}
where $a^{(\prime)}$ represent the combination of the Wilson
coefficients appearing in Tables~\ref{amp} and \ref{ampp}. The
Sudakov exponents associated with the various mesons are written
as
\begin{eqnarray}
S_{B}(t)&=&\exp\left[-s(x_{1}P_{1}^{+},b_{1})
-\frac{5}{3}\int_{1/b_{1}}^{t}\frac{d{\bar{\mu}}} {\bar{\mu}}
\gamma_q (\alpha _{s}({\bar{\mu}}))\right]\;,
\label{sb1} \\
S_{2}(t)&=&\exp\left[-s(x_{2}P_{2}^{+},b_2)
-s((1-x_{2})P_{2}^{+},b_{2})
-2\int_{1/b_{2}}^{t}\frac{d{\bar{\mu}}}{\bar{\mu}} \gamma_q
(\alpha_{s}({\bar{\mu}}))\right]\;, \label{sbk1}
\end{eqnarray}
with the quark anomalous dimension $\gamma_q=-\alpha_s/\pi$. The
formula for the exponential $S_3$ is the same as $S_2$ but with
the kinematic variables of meson 2 being replaced by those of
meson 3. The explicit expression of the exponent $s$ can be found
in \cite{LS,CS,BS}. The variable $b_{1}$, conjugate to the parton
transverse momentum $k_{1T}$, represents the transverse extent of
the $B$ meson. The transverse extents $b_2$ and $b_3$ have the
similar meaning for mesons 2 and 3, respectively. For the running
coupling constant $\alpha_{s}({\bar{\mu}})$, we employ the
one-loop expression, and the QCD scale $\Lambda_{\rm QCD}^{(4)}\,
=\, 0.250\;{\rm GeV}$. The Sudakov exponential decreases fast in
the large $b$ region, such that the long-distance contribution to
the decay amplitude is suppressed.

The hard functions are written as
\begin{eqnarray}
%---------------------------------------------------------------------
h_{e}(A,B,b_1,b_2,x_i) &=& \left[
  \theta\left( b_1-b_2 \right)
    K_0\left( A b_1 \right)   I_0\left( A b_2 \right)
\right.\nonumber\\
& &\left.\hspace{17mm}
  + \theta\left( b_2-b_1 \right)
    K_0\left( A b_2 \right)   I_0\left( A b_1 \right)
\right]
  K_0\left( B b_1 \right)\, S_t(x_i)
\;,\label{he}\\
%--------------------------------------------------
h_{n}(A,B,b_1,b_3) &=&
 K_0\left( A b_3 \right)
\left[
  \theta\left( b_1-b_3 \right)
    K_0\left( B b_1 \right)   I_0\left( B b_3 \right)
\right.\nonumber\\
& &\left.\hspace{40mm}
  + \theta\left( b_3-b_1 \right)
    K_0\left( B b_3 \right)   I_0\left( B b_1 \right)
\right] \;,
%--------------------------------------------------
\end{eqnarray}
where $S_t$ resums the threshold logarithm $\ln^2 x$ appearing the
hard kernels to all orders. It has been parameterized as \cite{UL}
\begin{eqnarray}
S_t(x)\, =\, \frac{2^{1+2c}\,\Gamma(3/2+c)}{\sqrt{\pi}\,\Gamma(1+c)}\,
[x(1-x)]^c\;,\label{str}
\end{eqnarray}
with $c= 0.3$.

The factorization formulas for ${\cal M}^{(q)}_{\pi K}$,
$q=u,c,t$, involve the convolutions of all three meson
distribution amplitudes:
\begin{eqnarray}
{\cal M}^{(q)}_{\pi K} &=& - 16  m_B^2 \frac{C_F^2}{\sqrt{2N_c}}
\int_0^1 dx_1 dx_2 dx_3\int_0^{\infty} b_1db_1 b_2db_2
\phi_B(x_1,b_1)
\nonumber \\
& & \times \big\{\left[
      (1+x_2) \phi_\pi^A(\overline{x_2})\phi_{K}^A(\overline{x_3})
        + r_\pi (1-2x_2)\left( \phi_\pi^P(\overline{x_2})-
\phi_\pi^T(\overline{x_2}) \right)
       \phi_{K}^A(\overline{x_3})
    \right. \nonumber \\
    & & \;\;\;\;\;\;\; \left.
    +2 r_K \phi_\pi^A(\overline{x_2}) \phi_{K}^P(\overline{x_3})
    +2 r_\pi r_K
       \left(
         (2+x_2)\phi_\pi^P(\overline{x_2})+x_2\phi_\pi^T(\overline{x_2})
       \right)  \phi_{K}^P(\overline{x_3})
\right]\nonumber \\
& & \;\;\;\;\;\;\;\;\times E^{(q)}(t_q,l^2) h_{e}(A,B,b_1,b_2,x_2)
\nonumber\\
& & \;\;\;\; + \left[
    2 r_\pi \phi_\pi^P(\overline{x_2})  \phi_{K}^A(\overline{x_3})
    +4 r_\pi r_K
       \phi_\pi^P(\overline{x_2})
       \phi_{K}^P(\overline{x_3})
\right]\nonumber\\
& &\;\;\;\;\;\;\;\;\times E^{(q)}(t_q^{\prime},l^{\prime 2})
h_{e}(A',B',b_2,b_1,x_1) \big\} \;, \label{ucp}
\end{eqnarray}
with the evolution factor,
\begin{eqnarray}
E^{(q)}(t,l^2)\, =\, \left[\alpha_s(t)\right]^2 C^{(q)}(t,l^2)
\exp\left[-S_B(t)-S_\pi(t)\right]\;.\label{ep}
\end{eqnarray}
The hard scales are chosen as
\begin{eqnarray}
t_q &=& {\rm max}\left(\sqrt{|A^2|},\,\sqrt{|B^2|},\,
\sqrt{\overline{x_2}x_3}m_B,\, 1/b_i \right) \;,
\nonumber\\
t_q^{\prime} &=& {\rm max}\left(\sqrt{|A'^2|},\,\sqrt{|B'^2|},\,
\sqrt{|x_3-x_1|}m_B,\, 1/b_i \right) \;,\label{tpp}
\end{eqnarray}
with the index $i=1,2$.
The additional scales $\sqrt{\overline{x_2}x_3}m_B$ and
$\sqrt{|x_3-x_1|}m_B$, compared to those appearing in
Eq.~(\ref{ttp}), come from the gluon invariant masses
$l^2=(1-x_2)x_3m_B^2$ and $l^{\prime 2}=(x_3-x_1)m_B^2$ in
Figs.~\ref{loops}(a) and \ref{loops}(b), respectively. The
formulas for ${\cal M}^{(u,c,t)}_{\pi\pi}$ are derived from
Eq.~(\ref{ucp}) by substituting the mass ratio $r_\pi$ for $r_K$,
and the distribution amplitudes $\phi_{\pi}^A$ and $\phi_{\pi}^P$
for $\phi_{K}^A$ and $\phi_{K}^P$, respectively.

The magnetic-penguin amplitude is written, for the $B\to \pi K$
modes, as \cite{MISHIMA03}
\begin{eqnarray}
{\cal M}^{(g)}_{\pi K} &=& 16 m_B^4  \frac{C_F^2}{\sqrt{2N_c}}
\int_0^1 dx_1dx_2dx_3 \int_0^{\infty} b_1db_1\, b_2db_2\,
b_3db_3\, \phi_B(x_1,b_1)
\nonumber \\
& &\times\, \Big\{ \left[
    - (1-x_2)
      \left\{ 2 \phi_\pi^A(\overline{x_2})
        + r_\pi \left( 3 \phi_\pi^P(\overline{x_2})-\phi_\pi^T(\overline{x_2})
 \right)
    \right. \right.   \nonumber \\
& &\hspace{10mm}\left. + r_\pi x_2 \left(
\phi_\pi^P(\overline{x_2}) + \phi_\pi^T(\overline{x_2}) \right)
    \right\} \phi_{K}(\overline{x_3})\nonumber\\
& &\hspace{10mm}
    -\, r_K (1+x_2) x_3 \phi_\pi^A(\overline{x_2})
      \left( 3 \phi_{K}^P(\overline{x_3}) +
 \phi_{K}^T(\overline{x_3}) \right)
    \nonumber \\
 & &\hspace{10mm}
    -\, r_K r_\pi(1-x_2)
        \left( \phi_\pi^P(\overline{x_2}) + \phi_\pi^T(\overline{x_2}) \right)
        \left( 3 \phi_{K}^P(\overline{x_3}) -
 \phi_{K}^T(\overline{x_3}) \right)
    \nonumber \\
 & &\hspace{10mm}\left.
    -\, r_K r_\pi x_3(1-2x_2)
        \left( \phi_\pi^P(\overline{x_2}) - \phi_\pi^T(\overline{x_2}) \right)
        \left( 3 \phi_{K}^P(\overline{x_3}) +
 \phi_{K}^T(\overline{x_3}) \right)
\right]\nonumber \\
& &\hspace{10mm}\times E_{g}(t_q) h_g(A,B,C,b_1,b_2,b_3,x_2)
\nonumber\\
& & - \left[ 4 r_\pi
 \phi_\pi^P(\overline{x_2})
         \phi_{K}(\overline{x_3})
     + 2 r_K r_\pi x_3 \phi_\pi^P(\overline{x_2})
         \left( 3 \phi_{K}^P(\overline{x_3}) +
 \phi_{K}^T(\overline{x_3}) \right)
\right]\nonumber \\
& &\hspace{10mm}\times E_{g}(t_q^{\prime})
h_g(A',B',C',b_2,b_1,b_3,x_1) \Big\} \;,
\end{eqnarray}
with the evolution factor $E_{g}(t)$,
\begin{eqnarray}
E_{g}(t)&=& [\alpha_s(t)]^2 C_{8g}^{\rm
eff}(t)\exp[-S_B(t)-S_K(t)-S_\pi(t)] \;.
\end{eqnarray}
Since the terms proportional to $r_Kr_\pi$ develop the end-point
singularities as the invariant mass of the gluon from $O_{8g}$
vanishes ($x_3\to 0$), we have kept the transverse momentum
$k_{3T}$. This is the reason the Sudakov factor associated with
the kaon appears. The hard function is given by
\begin{eqnarray}
\lefteqn{h_g(A,B,C,b_1,b_2,b_3,x_i)}\hspace{0mm}\nonumber\\
&=& - S_t(x_i)\, K_{0}\left( B b_1 \right)
  K_0  \left( C b_3 \right)
\int_0^{\frac{\pi}{2}} d\theta\,
  \tan\theta
  J_0\left( A b_1 \tan\theta \right)
  J_0\left( A b_2 \tan\theta \right)
  J_0\left( A b_3 \tan\theta \right)\;,
\end{eqnarray}
with the virtual quark and gluon invariant masses,
\begin{eqnarray}
& &A=\sqrt{x_2}m_B\;,\;\;\;\;B=B'=\sqrt{x_1x_2}m_B\;,\;\;\;\;C= i
\sqrt{(1-x_2)x_3}m_B\;,\nonumber\\
& &A'=\sqrt{x_1}m_B\;,\;\;\;\; C'= \sqrt{x_1-x_3}m_B\;.
\end{eqnarray}
The hard scales $t_q^{(\prime)}$ are the same as in
Eq.~(\ref{tpp}) with the index $i=1,2,3$.

At last, we present the factorization formula for the
nonfactorizable amplitude ${\cal M}_{e4}$ with the parton
transverse degrees of freedom in the kaon being neglected. This
formula is employed to justify the approximate equality of an
amplitude without the end-point singularity evaluated in collinear
and $k_T$ factorization theorems. Dropping the parton transverse
momentum $k_{3T}$, the corresponding change is to remove the
factor $m_B^2$ in Eq.~(\ref{me4}) and the integration $\int
b_3db_3$. We also replace the hard functions by \cite{CL0504}
\begin{eqnarray}
h_n(A^{(\prime)},B^{(\prime)},b_1,b_3)&\to&
\frac{m_B^2}{B^{(\prime)2}-A^{(\prime)2}} \left(
\begin{array}{cc}
 K_{0}\left(A^{(\prime)}b_{1}\right)-K_{0}\left(B^{(\prime)}b_{1}\right)
 &  \mbox{for $A^{(\prime)2} \geq 0$}  \\
 \frac{i\pi}{2} H_{0}^{(1)}\left(\sqrt{|A^{(\prime)2}|}b_{1}\right)
 -K_{0}\left(B^{(\prime)}b_{1}\right)
   & \mbox{for $A^{(\prime)2} < 0$}
  \end{array} \right)\;,
\label{hjd}
\end{eqnarray}
where $A^{(\prime)}$ and $B^{(\prime)}$ have been defined in
Table~\ref{tab:abcphik}. Without $k_{3T}$, the conjugate variable
$b_2$ in the Sudakov exponent $S_\pi(t)$ is set to $b_1$.

\end{document}